\documentclass[amsmath,amssymb,aps,pre,reprint, showpacs, superscriptaddress]{revtex4-1}
\usepackage[T1]{fontenc}
\usepackage{graphicx}
\usepackage{amsthm}
\usepackage{amsmath}
\usepackage{amssymb}
\usepackage{dsfont}
\usepackage{bm}
\usepackage{epstopdf}
\usepackage{hyperref}
\usepackage{color}

\usepackage{xparse}
\ExplSyntaxOn
\NewDocumentCommand{\mref}{m}{\quinn_mref:n {#1}}
\seq_new:N \l_quinn_mref_seq
\cs_new:Npn \quinn_mref:n #1
 {
  \seq_set_split:Nnn \l_quinn_mref_seq { , } { #1 }
  \seq_pop_right:NN \l_quinn_mref_seq \l_tmpa_tl
  ( 
  \seq_map_inline:Nn \l_quinn_mref_seq
    { \ref{##1},\nobreakspace } 
  \exp_args:NV \ref \l_tmpa_tl 
  ) 
 }
\ExplSyntaxOff

\begin{document}
\newcommand\mpi{\affiliation{Max Planck Institute for Mathematics in the Sciences, Inselstr. 22, D-04103 Leipzig, Germany}}
\newcommand\uleip{\affiliation{Institut f\"ur Theoretische Physik, Universit\"at Leipzig,  Postfach 100 920, D-04009 Leipzig, Germany}}
\newcommand\cha{\affiliation{ 
 Charles University, Faculty of Mathematics and Physics, 
 Department of Macromolecular Physics, 
 V Hole{\v s}ovi{\v c}k{\' a}ch 2, 
 CZ-180~00~Praha, Czech Republic 
}}

\title[]{Physically consistent numerical solver for time-dependent Fokker--Planck equations}
\author{Viktor Holubec}
\email{viktor.holubec@mff.cuni.cz}\uleip\cha
\author{Klaus Kroy}\uleip
\author{Stefano Steffenoni}\uleip\mpi
\date{\today} 
\begin{abstract}
We present a simple thermodynamically consistent method for solving time-dependent
Fokker--Planck equations (FPE) for over-damped stochastic processes,
also known as Smoluchowski equations. It yields both transition and
steady-state behavior and allows for computations of
moment-generating and large-deviation functions of observables defined
along stochastic trajectories, such as the fluctuating particle current, heat and work. The key strategy is to approximate the FPE by a Master equation with transition rates in configuration space that obey a local detailed balance condition for arbitrary discretization. Its time-dependent solution is obtained by a direct computation of the time-ordered exponential, representing the propagator of the FPE, by summing over all possible paths in the discretized space. The method thus not only preserves positivity and normalization of the solutions, but also yields a physically reasonable total entropy production, regardless of the discretization. To demonstrate the validity of the method and to exemplify its potential for applications, we compare it against Brownian-dynamics simulations of a heat engine based on an active Brownian particle trapped in a time-dependent quartic potential.
\end{abstract}


\maketitle

\section{Introduction}
\label{sec:Int}

Many natural phenomena exhibit a time-scale separation between ``slow'' and
``fast'' degrees of freedom. The variables varying slowly in space or
time can then be characterized by a self-contained coarse-grained
dynamics, which is --- for not too extreme coarse-graining --- perceptibly
perturbed by fluctuations arising from the noisy dynamics of the fast
variables. 

The Fokker--Planck equation (FPE) represents a most comprehensive
description of such time-separated phenomena \cite{Risken1996}. It predicts not only the
average dynamics of the slow variables but directly addresses, in a
technically manageable way, their complete probability distribution,
which includes the relevant information about the fluctuations of the slow degrees of freedom induced
by the fast ones. To achieve this, all of the slow
variables need to be resolved explicitly in a so-called Markovian
description, such that the remaining fast variables evolve without
perceptible memory of the past dynamics. 

Over the last century, the
FPE found applications in various scientific disciplines ranging from
physics and chemistry to biology and ecology and even economy and
finance \cite{Risken1996, Kampen1985, Friedrich2011, tra12,
  Slanina2013, pau10, dra02}.  Needless to say that it can only in
very few special cases be solved analytically, so that one usually has
to resort to analytical approximations, computer simulations and
numerical methods \cite{Risken1996, nar11, sep15, pic13, Holubec2014,
  shi18}. Both the Fokker--Planck formulation of stochastic dynamics
and efficient techniques for its numerical solution become
particularly relevant for situations far from equilibrium, where the
slow variables are, as a rule, found to exhibit non-Gaussian
characteristic fluctuations that contain a crucial part of the information about the system of interest.

For the physical interpretation of this information, it is moreover crucial to also evaluate functionals defined along individual trajectories of the underlying stochastic process, which is one of the main tasks of stochastic thermodynamics. Important examples of such functionals are fluctuating particle currents, and fluctuating heat and work in systems of Brownian particles, individual proteins, or living bacteria, which often operate under conditions far from equilibrium \cite{sek10,sei12,Speck2016}.

In this paper we describe a simple thermodynamically consistent matrix numerical method (MNM) for solving over-damped FPEs with time-dependent coefficients, also known as Smoluchowski equations. The method can resolve not only the transition and long-time behavior of probability distributions described by the FPE, but it is also naturally applicable to computations of moment generating-functions (MGFs) and large-deviation functions (LDFs) for various types of functionals defined along the trajectories of the stochastic process described by the FPE. This is achieved by a discretization that transforms the FPE into a Master equation with transition rates that obey a local detailed balance condition.
The time evolution of its solution is calculated from the time-ordered exponential, representing the discretized FPE-propagator, by summing over all possible paths in the discretized configuration space. The MNM thus addresses all
of the mentioned functions directly and gives physically reasonable results both from a probabilistic and from a thermodynamic point of view, for arbitrary discretization. Namely, the MNM is constructed to preserve the normalization and positivity of the initial distribution, and predict the correct entropy production of the discrete models emerging upon discretization, at arbitrary resolution.

Towards the end of the paper, we  test the MNM and illustrate its power by focusing on a specific example, namely a heat engine based on an over-damped active particle trapped in a time-dependent quartic potential and communicating with a heat bath at a time-dependent temperature. We investigate the dynamics of the particle and the fluctuations of work and heat exchanged with the bath, using both the proposed MNM and Brownian dynamics simulations (BD), checking that both methods give the same results.

\section{Principles of the MNM}
\label{sec:NDyn}

For pedagogical reasons, we introduce the MNM for a two-dimensional stochastic system, parametrized by coordinates $x$ and $y$. Although such a system can represent diffusion of an abstract object in an abstract energy landscape, we find it helpful to allude, in our description, to the intuitive paradigmatic example of an overdamped Brownian particle. Furthermore, we assume that the diffusion matrix $\mathcal{D}$ and the mobility matrix $\mu$ are diagonal: $\mathcal{D} = {\rm diag}(D_x, D_y)$ and $\mathcal{\mu} = {\rm diag}(\mu_x, \mu_y)$.
An extension to higher dimensions and off-diagonal matrices $\mathcal{D}$ and $\mathcal{\mu}$ is straightforward. The FPE for the probability density function (PDF) $\rho(x,y,t)$ to find the system at time $t$ in microstate $(x, y)$ is the parabolic partial differential equation
\begin{multline}
\partial_t \rho(x,y,t) = \mathcal{L}(x,y,t)\rho(x,y,t) = \\ \partial_{x}\left[\partial_{x} D_x - \mu_x  F_x \right] \rho(x,y,t)\\  +
  \partial_{y}\left[\partial_{y} D_y - \mu_y  F_y \right] \rho(x,y,t)
\label{eq:FPGeneral}
\end{multline}
with generally time-and position-dependent diffusion coefficients $D_x > 0$ and $D_y > 0$, mobilities $\mu_x$ and $\mu_y$ and forces $F_x$ and $F_y$ in the $x$ and $y$ directions, respectively. The force $\mathbf{F} = (F_x, F_y)$ does not need to be conservative, stemming from some potential $U = U(x,y,t)$, such that $\mathbf{F} = -\nabla U = - (\partial_{x}U,\partial_{y}U)$. Below, we show that the most general form of the FPE that can be solved using the MNM is Eq.~\eqref{eq:FPGeneral} with time-dependent but position-independent diffusion coefficients \eqref{eq:FPGeneralTRDB}. If one is willing to sacrifice the thermodynamic consistency of the MNM, its minimal modification moreover allows to solve Eq.~(\ref{eq:FPGeneral}) in full generality, i.e. with all the coefficients $D_x$, $D_y$, $\mu_x  F_x$ and $\mu_y  F_y$ time- and position-dependent.

The main idea, exploited in this paper, to solve the complicated time-dependent equation~(\ref{eq:FPGeneral}) is to approximate the underlying time-and-space continuous stochastic process by a time continuous hopping process in a discrete configuration space. To this end, we approximate the FPE~(\ref{eq:FPGeneral}) with the Fokker-Planck operator $\mathcal{L}(x,y,t)$ by a master equation with a transition rate matrix $\mathcal{R}$:
\begin{equation}
\partial_t \rho(x,y,t) = \mathcal{L}(x,y,t)\rho(x,y,t)
\rightarrow \dot{\mathbf{p}}(t) = \mathcal{R}(t) \mathbf{p}(t) \;.
\label{eq:vaugueFPE-Master}
\end{equation}
Here $\mathbf{p}(t)$ is the vector of probabilities of occupation of the individual discrete states which approximates the PDF $\rho(x,y,t)$, and $\dot{\mathbf{p}}(t)$ denotes its total time-derivative. In case of time-independent coefficients in $\mathcal{L}$, the Master equation is simply solved by matrix exponentiation of the constant rate matrix $\mathcal{R}$, namely $\mathbf{p}(t) = \exp\left[(t-t_0)\mathcal{R}\right]\mathbf{p}(t_0)$. In case of time-dependent coefficients, the strategy is to construct a piece-wise time-constant approximation $\tilde{\mathcal{R}}(t)$ to the time-dependent rate matrix $\mathcal{R}(t)$, solve the master equation in the time intervals where $\tilde{\mathcal{R}}(t)$ is constant using matrix exponentiation and, finally, employ the Markov property of the stochastic process to construct an approximate solution by concatenation, i.e., by multiplying the matrix exponentials. 

Simple variants of the MNM were already used by one of the authors to investigate several model systems \cite{rya16,hol17,Ornigotti2017}. The main merits of the present paper are twofold. First, we generalize the previously used method to FPEs with time-dependent coefficients and show how to calculate MGFs and LDFs for various functionals, in such a setting. Second, in the previous works \cite{rya16,hol17,Ornigotti2017} the MNM was always presented only as a minimal recipe in technical appendices. Here, we provide a comprehensive derivation and discussion of the method, including all its important aspects.

The following sections give a detailed description of the MNM.
First, in Sec.~\ref{sec:SpaceDiscretization}, we specify the discretization mesh used throughout the paper. The precise meaning of thermodynamic consistency and the transition rates obeying the local detailed balance condition are described in Secs.~\ref{sec:detailed_balance} and \ref{sec:transition_rates}. In Sec.~\ref{sec:BoundCond}, we discuss several boundary conditions which can be implemented with the method. In Sec.~\ref{sec:solution_of_ME}, we show how to solve the approximate Master equation. The long Sec.~\ref{sec:fluctuations} is devoted to computations of MGFs and LDFs for various functionals defined along the trajectories of the stochastic process described by the FPE. The general presentation of the MNM is closed by a discussion of several practical issues and of its computational efficiency compared to other methods, in Sec.~\ref{sec:performance_of_MNM}.
After that, in Sec.~\ref{sec:Exmlp}, we show how to apply the general theory by guiding the reader through a solved example: a heat engine consisting of an active particle trapped in a time-dependent quartic potential and communicating with a bath with time-dependent temperature. We conclude in Sec.~\ref{sec:Concl}. In Appendix~\ref{app:DFC}, we show why the (locally) detailed-balanced Master equation, which is at the heart of the MNM, can not be used for solving FPEs with position dependent diffusion coefficients and what modifications of the MNM are necessary in order to solve Eq.~\eqref{eq:FPGeneral} in full generality.

\subsection{Space discretization scheme}
\label{sec:SpaceDiscretization}

\begin{figure}[t!]
  \centering \includegraphics[width=1.0\linewidth]{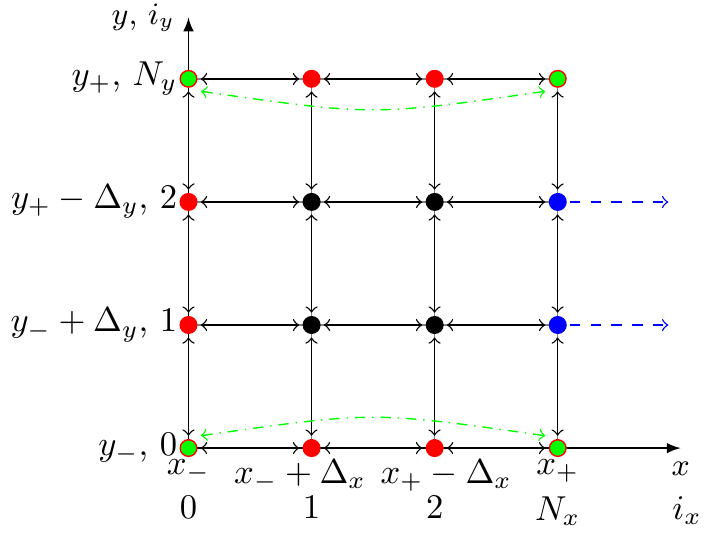}
  \caption{(Colored online) Sketch of the configuration space discretization used for the
    numerical solution of the two-dimensional overdamped Fokker-Planck
    equation~(\ref{eq:FPGeneral}). The black points mark the states inside the domain $[x_-,x_+]\times[y_-,y_+]$, the colored points form the boundary (see Sec.~\ref{sec:BoundCond}). The black full arrows depict the allowed transitions with the ``bulk'' transition rates \eqref{eq:ratexEQ} or \eqref{eq:ratexNEQ} (horizontal transition) and \eqref{eq:rateyEQ} or \eqref{eq:rateyNEQ} (vertical transitions). The red boundary is reflecting and thus the particles can not cross the red states (hence no red arrows). The blue boundary is absorbing and thus particles can leave the system from these sites (depicted by one-way dashed blue arrows). The states in the corners of the domain require two boundary conditions. In the figure, we impose reflecting boundary condition in the $y$ direction (depicted by red circumferences of the points) and periodic boundary conditions in the $x$ direction (depicted by green interiors of the points). The periodic boundary allows the particles to leave the system in the $x$ direction. The leaving particles reenter the system at the opposite side of the domain, as depicted by the green dot-dashed arrows.}
  \label{fig:space_discretization0}  
\end{figure}

Our goal is to solve the FPE Eq.~(\ref{eq:FPGeneral}) numerically. In general, this can be done only within some finite space-and-time domain, which allows to approximate the continuous space-time with a finite number of discrete points. For simplicity, we limit our presentation to rectangular domains of the form $[t_0,\tau]\times[x_-,x_+]\times[y_-,y_+]$ only. The generalization to more complicated domains is straightforward. The time domain is naturally bounded by the initial time $t_0$, where we impose an initial PDF $\rho(x,y,t_0)$, and the final time of integration $\tau$. The finite space domain $[x_-,x_+]\times[y_-,y_+]$ is defined by the boundary conditions imposed at boundaries $x = x_\pm$ and $y = y_\pm$. The boundary conditions which can be handled by the MNM will be detailed in Sec.~\ref{sec:BoundCond}. Here, we present the discretization of the (configuration) space domain $[x_-,x_+]\times[y_-,y_+]$ used in the rest of the paper.

For simplicity, we impose a rectangular discrete mesh with $(N_x + 1)(N_y + 1)$
discrete configurations with coordinates $\{i_x,i_y\}$,
\begin{eqnarray}
i_x &=& {\lfloor}\frac{x - x_-}{\Delta_x}{\rfloor}, \quad \Delta_x = \frac{x_+ - x_-}{N_x}\,,
\label{ix}\\
i_y &=& {\lfloor}\frac{y - y_-}{\Delta_y}{\rfloor}, \quad \Delta_y = \frac{y_+ - y_-}{N_y}\,,
\label{iy}
\end{eqnarray}
$i_x = 0,1,\dots,N_x$ and $i_y = 0,1,\dots,N_y$, as illustrated in Fig.~\ref{fig:space_discretization0}. The symbol ${\lfloor}x{\rfloor}$ denotes the floor function.
The generalization of the method to more complicated discretization meshes, which might be specifically adapted to some salient features of the coefficients in the FPE~\eqref{eq:FPGeneral}, is straightforward.

Let us denote as $p_{i_x,i_y} =
p_{i_x,i_y}(t)$ the occupation probabilities of the individual lattice points ${i_x,i_y}$. Allowing only transitions between neighboring lattice points (c.f.\ the arrows in Fig.~\ref{fig:space_discretization0}), the counterpart of the FPE~\eqref{eq:FPGeneral} on the discrete lattice is the Master equation \cite{Weber2017}
\begin{multline}
\dot{p}_{i_x,i_y} = r^{i_y}_{i_x+1 \to i_x}p_{i_x+1,i_y} + r^{i_y}_{i_x-1 \to i_x}p_{i_x-1,i_y} + \\ l^{i_x}_{i_y +1 \to i_y}p_{i_x,i_y+1} + l^{i_x}_{i_y -1 \to i_y}p_{i_x,i_y-1} - \\
\left(r^{i_y}_{i_x \to i_x + 1} + r^{i_y}_{i_x \to i_x - 1}
+ l^{i_x}_{i_y \to i_y + 1} + l^{i_x}_{i_y \to i_y
  - 1} \right)p_{i_x,i_y} \;,
\label{eq:MasterEQ}
\end{multline}
where the symbol $r^{i_y}_{i_x \to i_x + 1} = r^{i_y}_{i_x \to i_x + 1}(t) \ge 0$ denotes the transition rate in the $x$ direction from site $(i_x,i_y)$ to site $(i_x + 1,i_y)$ and $l^{i_x}_{i_y \to i_y + 1} = l^{i_x}_{i_y \to i_y + 1}(t) \ge 0$ denotes the transition rate in the $y$ direction from site $(i_x,i_y)$ to site $(i_x,i_y + 1)$. These transition rates must be chosen in such a way that the occupation probabilities $p_{i_x,i_y}$ determine the correct solution of the FPE~\eqref{eq:FPGeneral} in the limit of an infinitely fine mesh:
\begin{equation}
\rho(x,y,t) = \lim_{\Delta_x \to 0}\lim_{\Delta_y \to 0}\frac{p_{i_x(x),i_y(y)}(t)}{\Delta_x \Delta_y}\,.
\label{eq:mapping_rho_p}
\end{equation}
The Master Eq.~\eqref{eq:MasterEQ} possesses a simple probabilistic interpretation. For example, the expression $r^{i_y}_{i_x+1 \to i_x}(t)p_{i_x+1,i_y}(t)dt$ stands for the probability to jump from the site $(i_x+1,i_y)$ at time $t$ to the site $(i_x,i_y)$ during the infinitesimally short time interval $dt$. The time derivative of the occupation probability in Eq.~\eqref{eq:MasterEQ} is thus given by the probability to enter the site from neighboring sites [positive terms in \eqref{eq:MasterEQ}] minus the probability to leave it to neighboring sites [negative terms in \eqref{eq:MasterEQ}], during an infinitesimally short time interval. 

\subsection{Thermodynamic consistency}
\label{sec:detailed_balance}

The probabilistic interpretation of the Master Eq.~\eqref{eq:MasterEQ} implies that solutions produced by the proposed discretization are by construction non-negative for any non-negative initial condition and conserve the normalization in absence of source- or sink-boundary conditions (cf. Sec.~\ref{sec:BoundCond}), regardless of the dicretisation parameters $\Delta_x$ and $\Delta_y$. 

There are various ways how to write the rates for transitions between the lattice points depicted in Fig.~\ref{fig:space_discretization0} that lead to the same FPE~\eqref{eq:FPGeneral} in the limit of infinitely fine discretization. Here we want to propose a mapping~\eqref{eq:vaugueFPE-Master} guided by the aim to approximate the process described by the FPE~\eqref{eq:FPGeneral} in a thermodynamically consistent way, for arbitrary mesh resolution. A discretization scheme with similar properties was proposed already in 1970 by Chang and Cooper~\cite{Chang1970}. Compared to their presentation, our interpretation of the discretization scheme in terms of Master equations provides a clearer physical interpretation of the transition rates and a natural basis for studying various functionals, defined on realizations of the stochastic process, in terms of moment generating functions and characteristic functions.

On the level of coarse-grained stochastic models, the time reversal symmetry of the (standard) microscopic Hamiltonian dynamics manifests itself in a so-called \emph{local detailed balance condition} \cite{Jarzynski2000,mae13,Maes2003,Komatsu2008,Nakagawa2013}.
This condition should therefore be expected to hold for any physically reasonable stochastic dynamics. In fact, it can be viewed as the most fundamental tool for devising consistent thermodynamic notions for a microscopically grounded stochastic system. It states that the logarithm of ratio of the (conditional) path probability $P(\mathbf{r}_i \to \mathbf{r}_f,\mathbf{\Gamma}) = P(\mathbf{\Gamma})$ for the system to go from $\mathbf{r}_i$ to $\mathbf{r}_f$ along the path $\mathbf{\Gamma}$ over the probability $P^\star(\mathbf{r}_f\to \mathbf{r}_i,\mathbf{\Gamma}^\star) = P^\star(\mathbf{\Gamma}^\star)$ to return from $\mathbf{r}_f$ to $\mathbf{r}_i$ along the time-reversed path $\mathbf{\Gamma}^\star$ (with time-reversed dynamics) is proportional to the entropy change $\Delta S_{\rm R}(\mathbf{r}_i\to \mathbf{r}_f, \mathbf{\Gamma}) = \Delta S_{\rm R}(\mathbf{\Gamma})$ in the reservoir to which the system is connected along the path $\mathbf{\Gamma}$, briefly
\begin{equation}
k_{\rm B} \log \frac{P(\mathbf{\Gamma})}{P^\star(\mathbf{\Gamma}^\star)} = \Delta S_{\rm R}(\mathbf{\Gamma}).
\label{eq:detail_balance_general}
\end{equation}

Let us consider an overdamped diffusion process where a particle communicates with a single global equilibrium bath at constant temperature $T$ and is driven by a force $\mathbf{F} = (F_x,F_y)$. The fluctuation-dissipation theorem implies that the bath temperature is given by $T= D_x/(k_{\rm B} \mu_x)= D_y/(k_{\rm B} \mu_y)$ with time-and-space constant diffusion coefficients $D_x$ and $D_y$ and mobilities $\mu_x$ and $\mu_y$. The amount of entropy produced when the particle diffuses from $\mathbf{r}_i=(x_i,y_i)$ to $\mathbf{r}_f=(x_f, y_f)$ along the path $\mathbf{\Gamma} = [x(t),y(t)]$ parametrized by $t \in [t_i,t_f]$ is given by the energy transferred to the reservoir along this process divided by the reservoir temperature $T$. For overdamped dynamics, the energy dissipated into the bath is given by the work $W(\mathbf{\Gamma}) = \int_{\mathbf{\Gamma}} \mathbf{F}(\mathbf{\Gamma}) \cdot d\mathbf{\Gamma} = \int_{t_i}^{t_f}dt\, \mathbf{F}[x(t),y(t),t]\cdot[dx(t),dy(t)]/dt$ done by the force $\mathbf{F}$ along $\mathbf{\Gamma}$, and thus $\Delta S_{\rm R}(\mathbf{\Gamma}) = W(\mathbf{\Gamma})/T$. 

This equation can be generalized to situations where we connect the system at every point $(x,y)$ to one joint reservoir or even
two independent reservoirs with time-dependent temperatures. The bath at temperature $T_x(x,y,t) = D_x/(k_{\rm B} \mu_x)$ induces diffusion in the $x$-direction, the one with temperature $T_y(x,y,t) = D_y/(k_{\rm B} \mu_y)$ induces diffusion in the $y$-direction. Here, we again assumed that the diffusion and the mobility matrices $\mathcal{D}$ and $\mathcal{\mu}$ in Eq.~\eqref{eq:FPGeneral} are related by the fluctuation-dissipation theorem for each coordinate. In this generalized case the total amount of entropy produced in all the reservoirs along the trajectory $\Gamma$ reads 
\begin{multline}
\Delta S_{\rm R}(\mathbf{\Gamma}) = \int_{t_i}^{t_f}dt\, \left[\frac{F_x(t)}{T_x(t)},\frac{F_y(t)}{T_y(t)}\right]\cdot\left[\frac{dx(t)}{dt},\frac{dy(t)}{dt}\right] \\
= \int_{\mathbf{r}_i}^{\mathbf{r}_f}\left[\frac{F_x(t)}{T_x(t)},\frac{F_y(t)}{T_y(t)}\right]\cdot\left[dx(t),dy(t)\right]\;.
\label{eq:DeltaS_R_general}
\end{multline}

In order to find a reasonable form of the transition rates [transition probabilities $P(\mathbf{\Gamma})$ per unit time] based on Eqs.~\eqref{eq:detail_balance_general} and \eqref{eq:DeltaS_R_general}, we assume that the explicit time dependence of the forces and temperatures can be neglected for the transition rates at time $t_i$. If such a timescale separation holds, we can evaluate the force and temperature fields in Eq.~\eqref{eq:DeltaS_R_general} at time $t_i$, thereby effectively approximating the process with time-dependent coefficients by a sequence of processes with time-independent coefficients. For an over-damped diffusion process with time-independent coefficients the forward and reversed dynamics are identical, i.e. $P^\star(\mathbf{\Gamma})=P(\mathbf{\Gamma})$. Let us now use this formula to uncover the functional dependence of the transition probabilities, fulfilling the local detailed balance condition~\eqref{eq:detail_balance_general}, on the entropy change $\Delta S_{\rm R}\left(\mathbf{\Gamma}\right)$. 

Without loss of generality, we write the transition probability as 
$P(\mathbf{\Gamma}) = A(\mathbf{\Gamma})\exp\left[B\left(\mathbf{\Gamma}\right)/k_{\rm B}\right]$,
where $A$ denote a symmetric and $B$ an anti-symmetric unknown function with respect to the path reversal, i.e. $A(\mathbf{\Gamma}) = A(\mathbf{\Gamma}^\star)$ and $B(\mathbf{\Gamma}) = - B(\mathbf{\Gamma}^\star)$. Inserting this ansatz into Eq.~\eqref{eq:detail_balance_general} and using the condition $P^\star(\mathbf{\Gamma})=P(\mathbf{\Gamma})$, we find that $B(\mathbf{\Gamma}) - B(\mathbf{\Gamma}^\star) = 2B(\mathbf{\Gamma}) = \Delta S_{\rm R}\left(\mathbf{\Gamma}\right)$. We thus arrive at the expression
\begin{equation}
\frac{P(\mathbf{\Gamma})}{A(\mathbf{\Gamma})} = \frac{A(\mathbf{\Gamma})}{P(\mathbf{\Gamma}^\star)} = \exp{\left[\frac{\Delta S_{\rm R}(\mathbf{\Gamma})}{2 k_{\rm B}}\right]}
\label{eq:rates0}
\end{equation}
for the transition probabilities, the validity of which we assume for each transition and, consequently, also for arbitrary sequence of transitions. The prefactor $A(\mathbf{\Gamma})$ depends on the details of the dynamics and we determine it by inserting the transition rates fulfilling \eqref{eq:rates0} into the FPE.


\section{Implementation of the MNM}

\subsection{Transition rates}
\label{sec:transition_rates}

The formulas~\eqref{eq:rates0} can be applied to arbitrary discretization meshes. Let us now identify the points $\mathbf{r}_i$ and $\mathbf{r}_f$ with neighboring sites of the rectangular lattice defined in Sec.~\ref{sec:SpaceDiscretization} and depicted in Fig.~\ref{fig:space_discretization0}. We now take $\mathbf{r}_i=(x,y)$ and $\mathbf{r}_f = \mathbf{r}_f^x=(x + \Delta_x, y)$ for the horizontal transitions and $\mathbf{r}_f = \mathbf{r}_f^y =(x, y + \Delta_y)$ for the vertical ones. The probabilities $P(\mathbf{r}_i \to \mathbf{r}_f)$ will now determine the transition rates between the individual lattice points.

The formulas~\eqref{eq:rates0} imply that the necessary condition for the transition rates in the $x$ direction in Eq.~\eqref{eq:MasterEQ} to obey the local detailed balance principle is 
\begin{equation}
\frac{r_{i_x \to i_x + 1}}{A_{i_x + 1/2}} =
\frac{A_{i_x + 1/2}}{r_{i_x+1 \to i_x}}
= \exp{\left[\frac{\Delta S_{\rm R} (\mathbf{r}_i\to \mathbf{r}_f^x)}{2 k_{\rm B}}\right]},
\label{eq:xrates}
\end{equation}
where $A_{i_x + 1/2}$ is a symmetric prefactor, and similarly for the rates in the $y$ direction. In Appendix~\ref{app:DFC}, we show that the transition rates satisfying
these conditions can (even in one dimension) yield the FPE~\eqref{eq:FPGeneral} only for position-independent diffusion coefficients. Hence the most general FPE, which can be solved numerically using such transition rates reads
\begin{multline}
\partial_t \rho(x,y,t) = \mathcal{L}(x,y,t)\rho(x,y,t) = \\ \left[D_x \partial_{x}^2 - \partial_{x} \mu_x  F_x \right] \rho(x,y,t)\\  +
  \left[D_y \partial_{y}^2 - \partial_{y} \mu_y  F_y \right] \rho(x,y,t) \;.
\label{eq:FPGeneralTRDB}
\end{multline}
Nevertheless, in Appendix~\ref{app:DFC}, we also show how to modify the detailed-balanced transition rates in order to address the FPE~\eqref{eq:FPGeneral} in its full generality. The resulting generalized MNM respects the local detailed balance condition in case of position-independent diffusion coefficients. For position-dependent diffusion coefficients, the local detailed balance condition and the underlying microreversibility, valid in the continuous FPE~\eqref{eq:FPGeneral}, are thus necessarily broken on the coarse-grained level of the Master Eq.~\eqref{eq:MasterEQ}. This anticipates problems of attempts to mimic effects caused by spatially modulated mobilities using models with (temporally) diffusing diffusivities, see for example Ref.~\cite{Chechkin2017}.

\subsubsection{Equilibrium dynamics}
\label{sec:EQ_dyn}

Whenever the quantities $F_x/k_{\rm B} T_x = \mu_x F_x/D_x$ and $\mu_y  F_y/k_{\rm B} T_y = \mu_y F_y/D_y$ can be written using a dimensionless potential $\tilde{U}(x,y,t)$, such that 
\begin{equation}
\left(\frac{F_x}{k_{\rm B} T_x}, \frac{F_y}{k_{\rm B} T_y}\right)= -\nabla \tilde{U} = -(\partial_x \tilde{U}, \partial_y \tilde{U})\,,
\label{eq:Eq_Dym}
\end{equation}
the formula~\eqref{eq:DeltaS_R_general} can be written as $\Delta S_{\rm R}(\mathbf{\Gamma}) = \Delta S_{\rm R}(\mathbf{r}_i \to \mathbf{r}_f) = k_{\rm B}\left[\tilde{U}(\mathbf{r}_i,t) - \tilde{U}(\mathbf{r}_f,t)\right]$. The transition rates satisfying the condition \eqref{eq:rates0} and yielding the FPE~\eqref{eq:FPGeneralTRDB} in the limit $\Delta_x \to 0$, $\Delta_y \to 0$  of the Master Eq.~\eqref{eq:MasterEQ} can then be found without any further approximation by inserting the rates of the form \eqref{eq:xrates} into the FPE, similarly as in Appendix~\ref{app:DFC}. They read
\begin{eqnarray}
r^{i_y}_{i_x \to i_x \pm 1} &=& \frac{D_x}{\Delta_x^2}\exp{\left(-\frac{\tilde{U}_{i_x \pm 1,i_y} -
    \tilde{U}_{i_x,i_y}}{2}\right)}\;,
\label{eq:ratexEQ}\\
l^{i_x}_{i_y \to i_y \pm 1} &=&
\frac{D_y}{\Delta_y^2}\exp{\left(-\frac{\tilde{U}_{i_x,i_y \pm 1} -
    \tilde{U}_{i_x,i_y}}{2}\right)} \;,
\label{eq:rateyEQ}
\end{eqnarray}
with $\tilde{U}_{i_x,i_y} = \tilde{U}_{i_x,i_y}(t) = \tilde{U}(x_- + \Delta_x i_x, y_- + \Delta_y i_y,t)$ and the symmetric prefactors $D_x(t)/\Delta_x^2$ and $D_y(t)/\Delta_y^2$.

We refer to this as equilibrium dynamics because the FPE~\eqref{eq:FPGeneralTRDB} with time-independent coefficients fulfilling~\eqref{eq:Eq_Dym} leads to the Boltzmann stationary distribution $\rho(x,y,\infty) = \tilde{\rho}(x,y) \propto \exp\left[- \tilde{U}(x,y)\right]$. This can be verified by the direct substitution of the Boltzmann distribution into Eq.~\eqref{eq:FPGeneralTRDB}. Similarly, the stationary solution
of the Master Eq.~\eqref{eq:MasterEQ} reads $p_{i_x, i_y}(\infty) = \tilde{p}_{i_x i_y} \propto \exp\left(- \tilde{U}_{i_x i_y}\right)$, regardless of the discretization.

Physically, the most important feature of the equilibrium stationary distribution is that in this state all mesoscopic probability currents in the system vanish. On the level of the FPE~\eqref{eq:FPGeneralTRDB}, this is reflected by the formulas $j_x = - D_x\partial \tilde{\rho} + \mu_x F_x \tilde{\rho} = 0$ and $j_y = - D_y\partial \tilde{\rho} + \mu_y F_y \tilde{\rho} = 0$. On the level of the Master Eq.~\eqref{eq:MasterEQ}, the probability current in the $x$-direction reads $j_x^{i_y}(i_x \to i_x + 1) = r^{i_y}_{i_x \to i_x + 1} p_{i_x, i_y} - r^{i_y}_{i_x + 1 \to i_x} p_{i_x + 1, i_y}$ and similarly for the probability current in the $y$-direction. That these currents vanish for the Boltzmann distribution $\tilde{p}_{i_x i_y}$ is usually written in the form of the conventional \emph{global detailed balance conditions}
\begin{eqnarray}
\frac{r^{i_y}_{i_x \to i_x + 1}}{r^{i_y}_{i_x + 1 \to i_x}} &=&  \exp\left[- \left(\tilde{U}_{i_x + 1 i_y} - \tilde{U}_{i_x i_y}\right)\right]\;,
\label{eq:GDBEQx}\\
\frac{l^{i_x}_{i_y \to i_y + 1}}{l^{i_x}_{i_y + 1 \to i_y}} &=&  \exp\left[- \left(\tilde{U}_{i_x i_y + 1} - \tilde{U}_{i_x i_y}\right)\right]\;.
\label{eq:GDBEQy}
\end{eqnarray}
Let us stress that the ``equilibrium dynamics'' described in this section can sometimes be observed even though the system is not in equilibrium, for example, if the coefficients in the FPE~\eqref{eq:FPGeneralTRDB} are time-dependent and/or if the system relaxes from a non-equilibrium initial distribution $\rho \neq \tilde{\rho}$.

\subsubsection{Non-equilibrium dynamics}
\label{sec:NEQ_dyn}

If the quantities $F_x/k_{\rm B} T_x = \mu_x F_x/D_x$ and $\mu_y  F_y/k_{\rm B} T_y = \mu_y F_y/D_y$ can not be written using a single potential, one can still formally define (different) pseudo-potentials for the individual degrees of freedom:
\begin{equation}
\left(\frac{F_x}{k_{\rm B} T_x}, \frac{F_y}{k_{\rm B} T_y}\right)= -(\partial_x \tilde{U}, \partial_y \tilde{V})\,.
\label{eq:NEq_Dym}
\end{equation}
In this case, it is not possible to get rid of the path dependence of the integral in Eq.~\eqref{eq:DeltaS_R_general} as it was done for the equilibrium dynamics. Therefore, we now assume that for the transitions in the $x$-direction the entropy change can be well approximated by  
$\Delta S_{\rm R}(\mathbf{\Gamma}) = \Delta S_{\rm R}\left[(x,y) \to (x+\Delta_x,y)\right] = \tilde{U}(x,y,t) - \tilde{U}(x+\Delta x,y,t)$. This means that, from all possible paths $\mathbf{\Gamma}$ between the points $(x,y)$ and $(x+\Delta_x,y)$, we consider only the one with $y$-coordinate fixed at $y$. We use a similar approximation also for the $y$-direction. These approximations become exact in the limit of vanishing $\Delta_x$ and $\Delta_y$. The transition rates satisfying Eq.~\eqref{eq:rates0} under these approximations and leading to the FPE~\eqref{eq:MasterEQ} as the $\Delta_x \to 0$, $\Delta_y \to 0$ limit of the Master Eq.~\eqref{eq:MasterEQ} read
\begin{eqnarray}
r^{i_y}_{i_x \to i_x \pm 1} &=& \frac{D_x}{\Delta_x^2}\exp{\left(-\frac{\tilde{U}_{i_x \pm 1,i_y} -
    \tilde{U}_{i_x,i_y}}{2}\right)}\;,
\label{eq:ratexNEQ}\\
l^{i_x}_{i_y \to i_y \pm 1} &=&
\frac{D_y}{\Delta_y^2}\exp{\left(-\frac{\tilde{V}_{i_x,i_y \pm 1} -
    \tilde{V}_{i_x,i_y}}{2}\right)} \;,
\label{eq:rateyNEQ}
\end{eqnarray}
with $\tilde{U}_{i_x,i_y} = \tilde{U}_{i_x,i_y}(t) = \tilde{U}(x_- + \Delta_x i_x, y_- + \Delta_y i_y,t)$ and similarly for $\tilde{V}_{i_x,i_y}$. While, for nonzero $\Delta_x$ and $\Delta_y$, these transition rates satisfy the local detailed balance condition~\eqref{eq:rates0} for the FPE only approximately, they satisfy it exactly on the discrete lattice depicted in Fig.~\ref{fig:space_discretization0}, where the neighboring lattice points are interconnected exclusively by a single transition channel. On this discrete lattice, the process described by the rates~\eqref{eq:ratexNEQ} and \eqref{eq:rateyNEQ} is thus perfectly thermodynamically consistent, yielding the correct entropy produced along the individual transitions regardless of the discretization.

For the non-equilibrium dynamics, not only the time-dependent dynamics is in general unknown, but also the characterization of the stationary distribution, attained in case of time-independent coefficients in the FPE~\eqref{eq:FPGeneralTRDB}, is a non-trivial task. The presence of persevering probability currents in such steady states implies that there might be stationary transport of particles, energy, etc. Formally, the transition rates~\eqref{eq:ratexNEQ} and \eqref{eq:rateyNEQ} still obey a form reminiscent of the global detailed balance conditions~\eqref{eq:GDBEQx} and \eqref{eq:GDBEQy}, namely
\begin{eqnarray}
\frac{r^{i_y}_{i_x \to i_x + 1}}{r^{i_y}_{i_x + 1 \to i_x}} &=&  \exp\left[- \left(\tilde{U}_{i_x + 1 i_y} - \tilde{U}_{i_x i_y}\right)\right]\;,
\label{eq:GDBNEQx}\\
\frac{l^{i_x}_{i_y \to i_y + 1}}{l^{i_x}_{i_y + 1 \to i_y}} &=&  \exp\left[- \left(\tilde{V}_{i_x i_y + 1} - \tilde{V}_{i_x i_y}\right)\right]\;,
\label{eq:GDBNEQy}
\end{eqnarray}
but now with different potentials for the two degrees of freedom $x$ and $y$. Intuitively, each of these conditions is trying to draw the system into the Boltzmann equilibrium corresponding to its own potential $\tilde{U}$ or $\tilde{V}$, respectively. Globally, this competition leads to a non-equilibrium stationary state.

\subsection{Boundary conditions}
\label{sec:BoundCond}

The conditions at the boundaries of the configurational space domain $[x_-,x_+]\times[y_-,y_+]$ require some extra care and give rise to modifications of the transition rates presented in the previous section. Briefly, while the rates \eqref{eq:ratexEQ}--\eqref{eq:rateyEQ} and \eqref{eq:ratexNEQ}--\eqref{eq:rateyNEQ} are determined by the forces, temperatures and mobilities explicitly appearing in the dynamic operator in the FPE~\eqref{eq:FPGeneralTRDB}, this is not necessarily true for the rates at the boundaries. The probabilistic interpretation of the Master equation described below Eq.~\eqref{eq:mapping_rho_p} allows a convenient implementation of arbitrary boundary conditions, which are thus also easily introduced into the MNM. We now show how to implement three basic types of boundary conditions.

\begin{enumerate}
	\item \emph{Reflecting  boundary condition}: The particle can not cross the boundary. 
	\item \emph{Periodic boundary condition}: After crossing the boundary at one side of the domain the particle returns to it, usually at its other side.
	\item \emph{Absorbing boundary condition}: The particle is annihilated once it hits the boundary.
\end{enumerate}

While the reflecting and periodic boundary conditions lead to the overall conservation of probability (no particles can leave the system), the absorbing boundary conditions lead to depletion of the system due to particle losses at the boundary. Besides using these three types of boundary conditions, one can use arbitrary combinations thereof (with some probability the particles can be allowed to leave the system, or to appear at its other side, etc.). 

\subsubsection{Reflecting boundary conditions}

Physically, the reflecting boundary condition corresponds to an infinite potential barrier. Overcoming such a barrier requires an infinite amount of energy from the reservoir which corresponds to an infinite change of entropy in Eq.~\eqref{eq:rates0} or potential in Eqs.~\eqref{eq:ratexEQ}--\eqref{eq:rateyEQ} and \eqref{eq:ratexNEQ}--\eqref{eq:rateyNEQ}. The crossing rate across a reflecting barrier is thus 0, in accord with the rates Eqs.~\eqref{eq:ratexEQ}--\eqref{eq:rateyEQ} and \eqref{eq:ratexNEQ}--\eqref{eq:rateyNEQ}.

Let us, for example, consider the situation depicted in Fig.~\ref{fig:space_discretization0}, where the red points at the boundary obey reflecting boundary conditions. Specifically, we consider the point with coordinates $(1,0)$ (the second one in the last line). Realizing that the transitions over the reflecting barrier are not allowed and that this point has only a single boundary towards negative $i_y$, the Master Eq.~\eqref{eq:MasterEQ} for this point reads
\begin{multline}
\dot{p}_{1,0} = r^{0}_{2 \to 1}p_{2,0} + r^{0}_{0 \to 1}p_{00} + l^{1}_{1 \to 0}p_{1,1} - \\
\left(r^{0}_{1 \to 2} + r^{0}_{1 \to 0}
+ l^{1}_{0 \to 1} \right)p_{1,0} \;,
\label{eq:MasterEQ_RBC}
\end{multline}
Note that the transitions from $(1,0)$ to $(1,-1)$ and back occur with zero transition rate (and thus they do not show up in the equation). For other points with reflecting boundary, the master equation should be constructed in a similar manner.

\subsubsection{Periodic boundary conditions}

For periodic boundary conditions, the transitions rates are still given by Eqs.~\eqref{eq:ratexEQ}--\eqref{eq:rateyEQ} and \eqref{eq:ratexNEQ}--\eqref{eq:rateyNEQ}, one just needs to make the index periodic at the point where the periodic boundary condition is imposed. Consider for example the situation depicted in Fig.~\ref{fig:space_discretization0}, where the upper left and upper right points are connected by the periodic boundary in the $x$-direction. Then the rate to the right from the site $(N_x, N_y)$ leads to the site $(N_x + \Delta_x, N_y) = (0, N_y)$ and thus it reads
\begin{equation}
r^{N_y}_{N_x \to 0} = \frac{D_x}{\Delta_x^2}\exp{\left(-\frac{\tilde{U}_{N_x + \Delta_x,N_y} -
    \tilde{U}_{N_x,N_y}}{2}\right)}\;.
\label{eq:periodic_rate}
\end{equation}
In the expression for the transition rate, we used $\tilde{U}_{N_x + \Delta_x,N_y} - \tilde{U}_{N_x,N_y} = \int_{x_- + N_x \Delta_x}^{x_-+(N_x+1) \Delta_x} dx F_x/k_{\rm B}T_x$ instead of $\tilde{U}_{0,N_y} - \tilde{U}_{N_x,N_y}$, because, although the sites $(N_x + \Delta_x, N_y)$ and $(0, N_y)$ coincide, the pseudo-potential $\tilde{U}$ may be discontinuous at the boundary for a non-conservative force $F_x/k_{\rm B}T_x$. 

Considering that the site $(N_x, N_y)$ also possesses a reflecting boundary condition towards larger values of $i_y$, the corresponding Master equation reads
\begin{multline}
\dot{p}_{N_x, N_y} = r^{N_y}_{0 \to N_x}p_{0, N_y} + r^{N_y}_{N_x-1 \to N_x}p_{N_x-1, N_y} + \\ l^{N_x}_{N_y -1 \to N_y}p_{N_x, N_y-1} - \\
\left(r^{N_y}_{N_x \to 0} + r^{N_y}_{N_x \to N_x - 1}
+ l^{N_x}_{N_y \to N_y
  - 1} \right)p_{N_x, N_y} \;,
\label{eq:MasterEQ_PBC}
\end{multline}
Other transitions across periodic boundaries should be handled in a similar manner.

\subsubsection{Sources, sinks and absorbing boundaries}

Further examples are \emph{source/sink boundary conditions} meaning that particles can enter/leave the system across the boundary. They can be realized by connecting the boundary state to a particle reservoir. If the reservoir constantly feeds particles into the boundary state (the rate to go from the reservoir to the system is larger than the rate to go back), the boundary state behaves as a source. Vice versa, if the particles leave the boundary state towards the reservoir faster then they return, the boundary behaves as a sink.

The absorbing boundary condition represents a specific example of the sink condition with diverging rate to the reservoir and vanishing rate back. Physically, it corresponds to an infinitely deep potential cliff. When a particle hits such a boundary, it can be thought to release an infinite amount of energy that is dissipated to the bath, corresponding to a negatively infinite entropy change in Eq.~\eqref{eq:rates0} or an infinite change of the potential in Eqs.~\eqref{eq:ratexEQ}--\eqref{eq:rateyEQ} and \eqref{eq:ratexNEQ}--\eqref{eq:rateyNEQ}. Under such circumstances, the transition rates~\eqref{eq:ratexEQ}--\eqref{eq:rateyEQ} and \eqref{eq:ratexNEQ}--\eqref{eq:rateyNEQ} diverge. 

In order to avoid including such infinite rates in the master equation, we take as ``auxiliary'' boundary points those bulk points next to the actual boundary. The transition rates from the bulk into this auxiliary boundary and from it to all neighboring grid points are given by Eqs.~\eqref{eq:ratexEQ}--\eqref{eq:rateyEQ} or \eqref{eq:ratexNEQ}--\eqref{eq:rateyNEQ}, while the actual boundary points are assigned a vanishing back rate into the bulk. Consider for example the situation depicted in Fig.~\ref{fig:space_discretization0}, where the point ($N_x,2$) at the end of the second row from the top possesses an absorbing boundary in the $x$-direction. From the discussion above it follows that the corresponding Master equation reads
\begin{multline}
\dot{p}_{N_x, 2} = r^{2}_{N_x-1 \to N_x}p_{N_x-1, 2} +  l^{N_x}_{3 \to 2}p_{N_x, 3} + l^{N_x}_{1 \to 2}p_{N_x, 1} - \\
\left(r^{2}_{N_x \to N_x + 1} + r^{2}_{N_x \to N_x - 1}
+ l^{N_x}_{2 \to 3} + l^{N_x}_{2 \to 1} \right)p_{N_x, 2} \;,
\label{eq:MasterEQ_ABC}
\end{multline}
Here, the transition rate $r^{2}_{N_x \to N_x + 1}$ for transitions out of the system is given by Eqs.~\eqref{eq:ratexEQ}--\eqref{eq:rateyEQ} or \eqref{eq:ratexNEQ}--\eqref{eq:rateyNEQ}. Since we assume that the absorbing boundary in the continuous space described by the FPE is located at $x_+ + \Delta_x$, the pseudo-potentials $\tilde{U}_{N_x + 1,2}$ and $\tilde{V}_{N_x + 1,2}$ needed to evaluate the rates are well defined. Other transitions across absorbing boundaries should be handled in a similar manner.

\section{Solution of the Master equation}
\label{sec:solution_of_ME}

Having described the transition rates in the approximate Master Eq.~\eqref{eq:MasterEQ}, we will briefly explain how this equation can be solved in various situations. The key step always consists in rewriting the Master Eq.~\eqref{eq:MasterEQ} in the matrix form 
\begin{equation}
\dot{\mathbf{p}}(t) = {\mathcal R}(t) \mathbf{p}(t)\;,
\label{eq:MasterMatr}
\end{equation} 
where the $(N_x+1)(N_y + 1)\times (N_x+1)(N_y + 1)$ matrix ${\mathcal R}(t)$ contains the transition rates (\ref{eq:ratexEQ})--(\ref{eq:rateyEQ}) or (\ref{eq:ratexEQ})--(\ref{eq:rateyNEQ}) in such a way that Eqs.~(\ref{eq:MasterEQ}) and (\ref{eq:MasterMatr}) are equivalent. The elements of the $(N_x+1)(N_y + 1)$ dimensional vector $\mathbf{p}(t)$ are given by the occupation probabilities
$p_{i_x,i_y}(t)$ One possible construction is \cite{rya16}
\begin{equation}
\mathbf{p}(t) = (p_{0, 0}, \dots, p_{N_x, 0}, p_{0,1}, \dots,p_{N_x, 1},\dots,p_{N_x, N_y})^\top\;,
\label{eq:mapping}
\end{equation}
where $\top$ denotes the transposition. In this case, the probability $p_{i_x, i_y}$ is contained in the element $j(i_x,i_y) = i_y(N_x + 1) + i_x + 1$ of the vector $\mathbf{p}(t)$. The inverse transformation reads 
\begin{eqnarray}
i_x(j) &=& j - i_y(j)(N_x + 1) - 1\;,
\label{eq:transx}\\
i_y(j) &=& \left\lfloor (j - 1)/(N_x + 1) \right\rfloor\;.
\label{eq:transy}
\end{eqnarray}

The time dependence of the rate matrix ${\mathcal R}(t)$ comes directly from the time dependence of the coefficients $D_x$, $D_y$, $\mu_x$, $\mu_y$, $F_x$ and $F_y$ in the FPE~\eqref{eq:FPGeneralTRDB} appearing in the expressions for the transition rate. For the reflecting and periodic boundary conditions described in the preceding section, the matrix ${\mathcal R}(t)$ is stochastic ($\sum_i \left[{\mathcal R}(t)\right]_{ij}=0$) and thus Eq.~\eqref{eq:MasterMatr} conserves normalization of the probability vector $\mathbf{p}(t)$. All the following methods of solution for Eq.~\eqref{eq:MasterMatr} in diverse situations are based on basic algebraic manipulations involving the rate matrix.

\subsection{Time-independent coefficients}
\label{sec:time_independent}

Let us start with the simplest situation of time-constant
coefficients in the FPE~\eqref{eq:FPGeneralTRDB} which leads to a time-independent rate matrix, ${\mathcal R}(t) = {\mathcal R}$. In this case the Green's function (to which we also refer as the propagator throughout the text) for Eq.~(\ref{eq:MasterMatr}) is given by the matrix exponential
\begin{equation}
{\mathcal U}(t,t_0) = \exp[{\mathcal R}\,(t-t_0)]
\label{eq:Propagator}
\end{equation}
and thus the time evolution of the probability vector $\mathbf{p}(t)$ departing from the initial condition $\mathbf{p}(t_0)$ is given by
\begin{equation}
\mathbf{p}(t) = {\mathcal U}(t,t_0)\mathbf{p}(t_0)\;.
\label{eq:p(t)tIND}
\end{equation}
If the system state converges to a time-independent steady-state $\mathbf{p}_{\infty}$ at late times, this steady state can be either determined from Eq.~\eqref{eq:p(t)tIND} as $\mathbf{p}_{\infty} = \lim_{t\to \infty} \mathbf{p}(\infty)$, or, much more conveniently, as an eigenvector of the rate matrix corresponding to the eigenvalue 0:
\begin{equation}
\dot{\mathbf{p}}_{\infty} = {\mathcal R} \mathbf{p}_{\infty} = 0\;.
\label{eq:SStIND}
\end{equation}
Because only jumps between the neighboring sites are allowed (see Fig.~\ref{fig:space_discretization0}), the time-independent jump matrix ${\mathcal R}$ is sparse. Especially (but not solely) for the computation of the steady state vector $\mathbf{p}_{\infty}$ from Eq.~\eqref{eq:SStIND}
 one can benefit from fast numerical procedures for sparse matrices (see Sec.~\ref{sec:performance_of_MNM} for more details).

\subsection{Time-dependent coefficients}
\label{sec:time_dependent}

The ability to calculate the propagator ${\mathcal U}(t,t_0)$ for FPEs with time-constant coefficients eventually allows us to obtain the Green's function for Eq.~(\ref{eq:FPGeneralTRDB}) with arbitrary time-dependent
coefficients. We discretize the relevant time interval
$[t_0,t_0+\tau)$ into $N_t$ time slices
  of length $\Delta_t = \tau/N_t$. We assume that the driving can
  be approximated by appropriately chosen constants during all of these
  intervals and that it may change only step-wise from one interval to
  the next. In other words, we replace the actual
time-dependent coefficients $D_x$, $D_y$, $\mu_x$, $\mu_y$, $F_x$ and $F_y$ in Eq.~\eqref{eq:FPGeneralTRDB}
by their piece-wise constant approximations $\bar{D}_x(t) = D_x(t_0 + i_t \Delta_t)$, $i_t = {\lfloor}(t-t_0)/\Delta_t{\rfloor}$ and similarly for the other coefficients. The propagators for the individual time-intervals, during which the driving is constant, can be obtained using the procedure described above. Denoting by ${\mathcal U}_i$, $i \ge 1$, the propagator ${\mathcal U}[t_0 + (i+1)\Delta_t, t_0 + i\Delta_t] \equiv \exp\left[ \mathcal{R}(t_0 + i \Delta_t) \Delta_t \right]$ corresponding to the $i$th time-interval and by ${\mathcal U}_0 \equiv {\mathcal I}$ the unit matrix, we obtain the approximate Green's function under continuous driving, for arbitrary $t$, $t_0+\tau>t>t_0$, as
\begin{equation}
{\mathcal U}(t,t_0) = \lim_{\Delta_t \to 0} \prod_{i=0}^{i_t(t)}
{\mathcal U}_i \;.
\label{eq:GreenApprox}
\end{equation}
With this Green's function, the time evolution of the probability vector $\mathbf{p}(t)$ follows again from Eq.~\eqref{eq:p(t)tIND}.

Let us note that the presented discretization of time is just one of many possible choices. While we evaluate the time-dependent parameters at time $t'=t$ in order to compute the state of the system at time $t + \Delta_t$, one can also use values of the time-dependent parameters at any other time $t'$ in the interval $(t, t + \Delta_t)$. What value $t'$ suites best a specific situation depends on the relaxation time of the system. If it is long compared to $\Delta_t$, one should take $t' = t$. On the other hand, if the relaxation is fast compared to $\Delta_t$, one should rather take $t' = t + \Delta_t$.

\section{Functionals defined along the stochastic process}
\label{sec:fluctuations}

Besides computing the distribution $\rho(x,y,t)$ and then using it to evaluate averages, moments, reduced distribution functions for $x$ and $y$, and the mesoscopic probability currents $j_x$ and $j_y$, the probabilistic interpretation of the discrete approximation~\eqref{eq:MasterEQ} of the FPE~\eqref{eq:FPGeneralTRDB} can moreover be employed to address the statistics of various stochastic variables, other than position, directly. Useful examples are microscopic currents or linear combinations thereof, and heat, work, or efficiency which are much studied objects in stochastic thermodynamics. 

Application of the MNM to probability currents was already suggested in Refs.~\cite{rya16,hol17}, where it was employed in the calculation of the diffusion coefficient in a model of a two-dimensional Brownian ratchet. Here we discuss this approach in greater generality.

\subsection{Probability currents}

The probability current $\mathbf{j}(x,y,t) = (j_x,j_y)$ at time $t$ and position $\mathbf{r} = (x,y)$ can be defined in two equivalent ways. First, one can define it \emph{mesoscopically}, rewriting the FPE~\eqref{eq:FPGeneralTRDB} as $\partial_t \rho(x,y,t) = {\mathcal L}(x,y,t) = - \nabla \cdot \mathbf{j}(x,y,t)$ leading to the expression
\begin{equation}
\mathbf{j}(x,y,t) = - \left(D_x\partial_x + \mu_x F_x, 
D_y\partial_y + \mu_y F_y \right) \rho \;.
\label{eq:cuurentFP}
\end{equation}
On the level of the Master Eq.~\eqref{eq:MasterEQ}, these expressions read 
\begin{multline}
\dot{p}_{i_x, i_y}(t) = j_x^{i_y}(i_x+1 \to i_x) + j_x^{i_y}(i_x - 1 \to i_x) + \\
j_y^{i_x}(i_y + 1 \to i_y) + j_y^{i_x}(i_y - 1 \to i_y)
\label{eq:micro_current_ME}
\end{multline}
and
\begin{eqnarray}
\!\!\!\!\!\! j_x^{i_y}(i_x \to i_x + 1) &=& r^{i_y}_{i_x \to i_x + 1} p_{i_x, i_y} - r^{i_y}_{i_x + 1 \to i_x} p_{i_x + 1, i_y}
\\
\!\!\!\!\!\! j_y^{i_x}(i_y \to i_y + 1) &=& l^{i_x}_{i_y \to i_y + 1} p_{i_x, i_y} - l^{i_x}_{i_y + 1 \to i_y} p_{i_x, i_y + 1}
\end{eqnarray}
The mappings between the probability currents in the continuous space and those on the discrete lattice read
\begin{eqnarray}
j_x(x,y,t) &=& \lim_{\Delta_x \to 0, \Delta_y \to 0} \frac{j_x^{i_y}(i_x \to i_x + 1,t)}{\Delta_y} \;,
\label{eq:current_ME1}
\\
j_y(x,y,t) &=& \lim_{\Delta_x \to 0, \Delta_y \to 0} \frac{j_y^{i_x}(i_y \to i_y + 1,t)}{\Delta_x}\;,
\label{eq:current_ME2}
\end{eqnarray}
where $x= x_- + \Delta_x i_x$ and $y = y_- + \Delta_y i_y$. The appearance of the factors $\Delta_x$ and $\Delta_y$ follows from discretization of the formula $\partial_t \rho = -\nabla\cdot\mathbf{j} = \dot{p}_{i_x,i_y}/{\Delta_x \Delta_y} = \sum j/{\Delta_x \Delta_y}$, valid in the limit of infinitely fine mesh, where $\sum j$ stands for right-hand side of Eq.~\eqref{eq:micro_current_ME}.

\emph{Microscopically}, the current can be defined as $\mathbf{j}(x,y,t) = \left< \delta[\mathbf{r}(t) - \mathbf{r}]\, \dot{\mathbf{r}}(t) \right> = \left< \delta[x(t) - x]\delta[y(t) - y]\, \dot{\mathbf{r}}(t) \right>$, where the average is taken over many trajectories $\mathbf{r}(t)$ of the underlying stochastic process. The quantity
\begin{equation}
\mathfrak{J}(x,y,t) = \mathfrak{J}(\mathbf{r},t) = \delta[\mathbf{r}(t) - \mathbf{r}]\, \dot{\mathbf{r}}(t)
\label{eq:micro_current}
\end{equation}
inside the average is what we call a \emph{microscopic}  current. In measurements, one can obtain not only the average current $\mathbf{j}$, but its full probability distribution. The MNM can be applied to investigate this distribution as well as other distributions of arbitrary variables that arise as linear combinations of the microscopic currents $\mathfrak{J}(x,y)$ at different positions. An important example of such a variable from the field of stochastic thermodynamics is heat, as exemplified in the example in Sec.~\ref{sec:Exmlp}.

The lattice equivalents of the microscopic definitions of the mesoscopic currents are the formulas $j_x^{i_y}(i_x \to i_x + 1) = \left<\frac{d i_x(t)}{t} \delta_{i_x(t) i_x}\delta_{i_y(t) i_y}   \right>$ and $j_y^{i_x}(i_y \to i_y + 1) = \left<\frac{d i_y(t)}{t} \delta_{i_x(t) i_x}\delta_{i_y(t) i_y}   \right>$. The $x$-current measures the number of jumps to the right from the lattice point minus the number of jumps from the right to the lattice point, and similarly for the $y$ current. 

\subsection{Moment generating functions for observables proportional to integrated currents}
\label{sec:MGF_current}

In this section, we calculate the moment generating function $\chi_A$
for an observable which is given by an arbitrary linear combination of the microscopic currents \eqref{eq:micro_current}
\begin{multline}
A(t_0+\tau,t_0) = \int_{t_0}^{t_0+\tau}dt\int dx \int dy\,
\mathbf{c}(\mathbf{r},t)\cdot\mathfrak{J}(\mathbf{r},t)\\
= \int_{t_0}^{t_0+\tau}dt\,
\mathbf{c}[\mathbf{r}(t),t]\cdot\dot{\mathbf{r}}(t)\;,
\label{eq:gen_current_variable}
\end{multline}
where $\mathbf{c}(\mathbf{r},t) = (\partial_x g, \partial_y h)$ is a vector of space- and time-dependent coefficients. The MGF $\chi_A = \int_{-\infty}^\infty dA\, \exp\left(-s_A A \right)p(A)$  is defined as a two-sided Laplace transform of the probability distribution $p(A)$. 

In Appendix~\ref{app:MGF_intergated _current} we discuss in detail the computation of the MGF $\chi_{\bar{\mathfrak{J}}}$ for the time-averaged probability current $\bar{\mathfrak{J}}(\mathbf{r},\tau) = \frac{1}{\tau}\int_{t_0}^{t_0+\tau}dt \mathfrak{J}(\mathbf{r},t)$. The MGF $\chi_A$ can be computed along similar lines as $\chi_{\bar{\mathfrak{J}}}$ and thus we here omit the details and present the main results only. 

The key ingredient in the computation of the MGF is the construction of the so-called tilted matrix $\tilde{\mathcal{R}}_{s_A}(t)$. In the present case, the rate matrix $\mathcal{R}(t)$ must be tilted  proportionally to the coefficients $\partial_x g$ and $\partial_y h$ in the vector $\mathbf{c}(\mathbf{r},t)$. Namely, the rates $r^{i_{y}}_{i_{x}\to i_{x} + 1}(t)$ must be multiplied by 
\begin{equation}
\exp\left\{-s_A \left[g(x_- + (i_{x}+1)\Delta_x,t) - g(x_- + i_{x}\Delta_x,t)\right]\right\}\;,
\end{equation}
 the rates $r^{i_{x}}_{i_{y} + 1\to i_{y}}(t)$ by 
\begin{equation}
\exp\left\{-s_A \left[h(y_- + i_{y}\Delta_y,t) - h(y_- + (i_{y} + 1)\Delta_y,t)\right]\right\}\;,
\end{equation}
and similarly for all other transition rates. The MGF for $A(t_0+\tau,t_0)$ is then obtained from Eq.~\eqref{eq:char_fun_n} with the only difference that the tilted matrices $\tilde{\mathcal{R}}_{\mathbf{s}_{\mathbf{n}}}(t)$ involved in the equation are substituted by the tilted matrices $\tilde{\mathcal{R}}_{s_A}(t)$ just described above. Namely,
\begin{equation}
\chi_A(s_A,t,t_0) = \lim_{\Delta_t \to 0} \mathbf{p}^\top_+  \prod_{i=0}^{i_t(t)}
\tilde{{\mathcal U}}_i(s_A)\mathbf{p}(t_0)\;,
\label{eq:MGF_A}
\end{equation}
where $\mathbf{p}^\top_+$ is a vector of ones effecting the summation over the final states at time $t=t_0+\tau$, and $\tilde{{\mathcal U}}_i(s_A) = \exp\left[\tilde{\mathcal{R}}_{s_A}(t_0 + i\Delta_t)\Delta_t \right]$ if $i>1$ and the unit matrix $\mathcal{I}$ otherwise. For problems with a time-independent tilted rate matrix $\tilde{\mathcal{R}}_{s_A}$, the product in Eq.~\eqref{eq:MGF_A} simplifies to $\prod_{i=0}^{i_t(t)} \tilde{{\mathcal U}}_i(s_A) = \exp\left[\tilde{\mathcal{R}}_{s_A} \tau \right] = \tilde{{\mathcal U}}(s_A,t_0+\tau,t_0)$ and the moment generating function is thus given by
\begin{equation}
\chi_A(s_A,t,t_0) =  \mathbf{p}^\top_+
\tilde{{\mathcal U}}(s_A,t,t_0)\mathbf{p}(t_0)\;.
\label{eq:MGF_TIA}
\end{equation}

Some examples of physically relevant observables of the type \eqref{eq:gen_current_variable} are time-averaged probability currents $j_x = \int dt \int dx \int dy\, \mathfrak{J}_x(x,y,t)/\tau$ flowing through the system in the $x$ direction [here $\tau \mathbf{c}(x,y,t) = (1,0)/\tau$]; time-averaged probability currents $j_y = \int dt \int dx \int dy\, \mathfrak{J}_y(x,y,t)$ flowing through the system in the $y$ direction [here $\tau \mathbf{c}(x,y,t) = (0,1)$]; the total heat flux $Q = \int dt \int dx \int dy\, \nabla U(x,y,t)\cdot\mathfrak{J}(x,y,t)/\tau$ flowing from the reservoirs into a Brownian ratchet \cite{rya16,hol17} [here $\tau \mathbf{c}(x,y,t) = \nabla U(x,t)$, where $U(x,y)$ is a potential energy]; and the heat flux $Q_x = \int dt \int dx \int dy\, \partial_x U(x,y,t)\cdot\mathfrak{J}(x,y,t)/\tau$ flowing into the ratchet from the reservoirs connected to the $x$ coordinate only [here $\tau \mathbf{c}(x,y,t) = (\partial_x U(x,t),0)$]. 

For the observables $A$ where the scalar product $\mathbf{c}(\mathbf{r},t)\cdot\mathfrak{J}(\mathbf{r},t)$ in Eq.~\eqref{eq:gen_current_variable} can be written in the form of a
total time derivative $d f [x(t),y(t),t]/dt = \partial f/\partial t + \nabla f \cdot \dot{\mathbf{r}}$,
the formula \eqref{eq:gen_current_variable} can be simplified as
\begin{multline}
A(t_0+\tau,t_0) = \int_{t_0}^{t_0+\tau}dt\, \frac{d}{dt}f[\mathbf{r}(t),t]\\
= f[\mathbf{r}(t_0+\tau),t_0+\tau] - f[\mathbf{r}(t_0),t_0]
\label{eq:gen_current_variable2}
\end{multline}
and thus depends only on the initial and final times and positions. Also in this case, the calculation of the MGF for $A$ can be simplified as in the step from Eq.~\eqref{eq:MGF_A} to Eq.~\eqref{eq:MGF_TIA}. Now, the matrix $\tilde{{\mathcal U}}(s_A,t_0+\tau,t_0) = \tilde{{\mathcal U}}(s_A)$ has elements
\begin{equation}
\left[\tilde{{\mathcal U}}(s_A)\right]_{kl} = \left[{\mathcal U}(t_0+\tau,t_0)\right]_{kl}{\rm e}^{- s_A\Delta_c(k,l,t_0+\tau,t_0)}\;,
\label{eq:UAti}
\end{equation}
where $\Delta_c(k,l,t_0+\tau,t_0) = f[\mathbf{r}_f,t_0+\tau] - f[\mathbf{r}_i,t_0]$, $\mathbf{r}_f = [x_- + \Delta_x i_x(k),y_- + \Delta_y i_y(k)]$, $\mathbf{r}_i = [x_- + \Delta_x i_x(l),y_- + \Delta_y i_y(l)]$. Here, the coefficients $i_x(k)$ and $i_y(k)$ are given by Eqs.~\eqref{eq:transx}--\eqref{eq:transy}. A typical example of such an observable is the above mentioned heat in case the potential $U(x,t)$ does not depend on $t$ explicitly. However, since we treat time-dependent protocols using a piece-wise constant approximation (see Sec.~\ref{sec:time_dependent}), this simplification is important also for time-dependent potentials. If the product $\mathbf{c}(\mathbf{r},t)\cdot\mathfrak{J}(\mathbf{r},t)$ can be written as a total derivative $df/dt$ only for a time-independent vector $\mathbf{c}(\mathbf{r},t) = \mathbf{c}(\mathbf{r})$, the moment generating function for $A$ with the explicitly time-dependent $\mathbf{c}(\mathbf{r},t)$ can be calculated from Eq.~\eqref{eq:MGF_A} with $\tilde{{\mathcal U}}_i(s_A)$, $i>1$, redefined using Eq.~\eqref{eq:UAti} as $\tilde{{\mathcal U}}[s_A,t_0 + (i+1)\Delta_t, t_0 + i\Delta_t]$.

\subsection{Moment generating functions for observables not proportional to integrated currents}

Above, we have focused solely on observables which can be written as linear combinations~\eqref{eq:gen_current_variable} of microscopic probability currents. The integrand in these observables vanishes if the particle does not move ($\dot{\mathbf r} = 0$). However, in driven systems, there are also important observables with nonzero increments even if the particle stands still. The MNM can also be used to calculate MGFs and LDFs for observables of the form
\begin{multline}
B(t_0+\tau,t_0) = \int_{t_0}^{t_0+\tau}dt\int dx \int dy\, \delta[\mathbf{r}(t) - \mathbf{r}]\partial_{t}b(\mathbf{r},t) \\
= \int_{t_0}^{t_0+\tau}dt\, \partial_{t}b[\mathbf{r}(t),t]
\;.
\label{eq:gen_noncurrent_variable}
\end{multline}
The observable $B$ vanishes if the function $b$ is constant in time. The best known example of a physically relevant observable of the type~\eqref{eq:gen_noncurrent_variable} is the stochastic work done on the system due to a deterministic external driving, which changes the potential $U$. Then we have $b[\mathbf{r}(t),t] = b[x(t),y(t),t] = U(x(t),y(t),t)$. Another example is the occupation time for a position $\mathbf{r}_a$, in which case $b(\mathbf{r}(t),t) = \delta\left[\mathbf{r}(t) - \mathbf{r}_a\right]t$, or the occupation time for a region $\Omega$, in which case $b(\mathbf{r}(t),t) = I_\Omega[\mathbf{r}(t)]t$, where $I_\Omega(\mathbf{r})$ is an indicator function equal to one if $\mathbf{r}\in \Omega$ and $0$ otherwise.

For observables of the above type $B$, the tilted matrix must be constructed using the time discretization, already introduced to derive Eq.~\eqref{eq:GreenApprox}. We define the piece-wise constant approximation of the function $b$ as $\bar{b}(\mathbf{r},t) = b(\mathbf{r},t_0 + \Delta_t i_t)$, $i_t = \lfloor (t-t_0)/\Delta_t \rfloor$. For this approximate function, the variable $B$ in Eq.~\eqref{eq:gen_noncurrent_variable} does not change during the time intervals $[t_0 + \Delta_t i, t_0 + \Delta_t (i + 1)]$, where $\bar{b}(\mathbf{r},t)$ is constant for constant $\mathbf{r}$, and it abruptly jumps from $B(t)$ to $B(t) + b[\mathbf{r}(t),t+] - b[\mathbf{r}(t),t-]$ at time instants $t = t_0 + \Delta_t i$, where $\bar{b}(\mathbf{r},t)$ changes infinitely fast. Here $b[\mathbf{r}(t),t\pm] = \lim_{\epsilon \to 0} b[\mathbf{r}(t),t\pm\epsilon]$, $\epsilon\ge 0$. 

Let us now turn to the discrete approximation of the full process using the discrete lattice of Fig.~\ref{fig:space_discretization0}. Using the notation of Eq.~\eqref{eq:GreenApprox} and assuming that the system is in microstate $[i_x(l),i_y(l)]$ at time $t_0 + \Delta_t i$ and in microstate $[i_x(k),i_y(k)]$ at time $t_0 + \Delta_t (i+1)$ [see Eqs.~\eqref{eq:transx} and \eqref{eq:transy} for definitions of $i_x(l)$ and $i_y(l)$], the PDF for $B$ is given by
\begin{equation}
\left[\bar{{\mathcal U}}_i(B)\right]_{kl} = \left[{\mathcal U}_i\right]_{kl}\delta\left[B - \Delta_b(k,t_0 + \Delta_t i)\right]\;,\; i \ge 1\;.
\label{eq:barUB}
\end{equation}
 Here we used the shorthand $\Delta_b(k,t) = b[\mathbf{r},t+]  - b[\mathbf{r},t-]$, $\mathbf{r} = [x_- + \Delta_x i_x(k),y_- + \Delta_y i_y(k)]$ and ${\mathcal U}_0(B) = {\mathcal I}$. The PDF for $B$ during the whole time interval $[t_0,t_0+\tau]$ is thus given by a multiple convolution of the form $\lim_{\Delta_t \to 0} \mathbf{p}^\top_+  [\bar{{\mathcal U}}_{i(t)}\star\bar{{\mathcal U}}_{i(t)-1} \star \dots \star\bar{{\mathcal U}}_0](B)\mathbf{p}(t_0)$. The MGF for $B$ and thus also the corresponding tilted matrix is obtained by the Laplace transform of the last expression with respect to $B$:
\begin{equation}
\chi_B(s_B,t_0+\tau,t_0) = \lim_{\Delta_t \to 0} \mathbf{p}^\top_+  \prod_{i=0}^{i_t(t_0+\tau)}
\tilde{{\mathcal U}}_i(s_B)\mathbf{p}(t_0)\;,
\label{eq:MGF_B}
\end{equation}
where the matrix $\tilde{{\mathcal U}}_i(s_B)$ is obtained as the Laplace transform of the matrix $\bar{{\mathcal U}}_i(B)$ (we again just substitute the $\delta$-functions $\delta[B - \Delta_b(k,t_0 + \Delta_t i)]$ for exponentials $\exp[-s_B \Delta_b(k,t_0 + \Delta_t i)]$).


The MNM can also be applied to variables which are defined as combinations of the variables of the types $A$ and $B$. An example of such a variable is the increase of internal energy $\Delta U = W + Q$, which consists of heat $Q$ (type $A$ variable) and work $W$ (type $B$ variable). Let us consider a general variable $C$ decomposed as $C = A + B$. Then the corresponding MGF $\chi_C$ is given by Eq.~\eqref{eq:MGF_B} with the tilted matrices $\tilde{\mathcal U}_i(s_C)$ given by
\begin{equation}
\left[\tilde{\mathcal U}_i(s_C)\right]
= [\tilde{\mathcal B}_i(s_C)]_{kl} \exp\left[
- s_C \Delta_b (k, t_0 + \Delta_t i)
\right]\;,
\end{equation}
where $\tilde{\mathcal B}_i(s_C)$ is the tilted matrix $\tilde{\mathcal U}_i$ for $A$, defined in Eq.~\eqref{eq:MGF_A}.
Similarly to the case of the variables of type $A$, also the computation of $\chi_C$ may simplify if the variable $C$ has a suitable structure.

\subsection{Moments and cumulants}
\label{sec:Moments_Cumulants}

The MGF $\chi_{X}(s,t,t_0)$ allows one to access all moments of the stochastic variable $X$ at time $t$ simply by taking derivatives with respect to the Laplace variable $s$:
\begin{equation}
\left<X^n(t)\right> = (-1)^n \frac{d^n \chi_{X}(s,t,t_0)}{d s^n}\bigg|_{s=0}\;.
\label{eq:moments_from_MGF}
\end{equation}
The zeroth moment is just a normalization $\chi_{X}(0,t,t_0) = 1$ and it can be used as a first test of the calculated MGF. The first moment equals the average $\left<X(t)\right>$ of the quantity $X$ and it can be calculated from the probability distribution for position $\rho(x,y,t)$ [or from its approximation $\mathbf{p}(t)$]. For the variable $A$ defined in Eq.~\eqref{eq:gen_current_variable} it reads
\begin{equation}
\left<A(t)\right> = \int_{t_0}^{t}dt'\int dx \int dy\,
\mathbf{c}(x,y,t')\cdot\mathbf{j}(x,y,t')\;,
\label{eq:first_moment_A}
\end{equation}
where the average current $\mathbf{j}(x,y,t)$ is given either by Eq.~\eqref{eq:cuurentFP} or by Eqs.~\eqref{eq:current_ME1} and \eqref{eq:current_ME2}. For the variable $B$ defined in Eq.~\eqref{eq:gen_noncurrent_variable} we get
\begin{equation}
\left<B(t)\right> = \int_{t_0}^{t}dt'\int dx \int dy\,
\partial_{t}b[x,y,t'] \mathbf{\rho}(x,y,t')\;.
\label{eq:first_moment_B}
\end{equation}
The formulas \eqref{eq:first_moment_A} and \eqref{eq:first_moment_B} can be used as another test of calculated MGFs.

In a similar manner to moments, the MGF can be used for calculating all cumulants $C_n(X,t)$ of the variable $X$ at time $t$:
\begin{equation}
C_n(X,t) = (-1)^n \frac{d^n \log \chi_{X}(s,t,t_0)}{d s^n}\bigg|_{s=0}\;.
\label{eq:cumulants_from_MGF}
\end{equation}
The cumulants reflect the shape of the probability distribution for $X$. First four of them can be written in terms of moment as $C_0 = 0$, $C_1 = \left<X\right>$, $C_2 = \left<X^2\right> - \left<X\right>^2$ and $C_3 = \left<X^3\right> - 3\left<X^2\right>\left<X\right> + 2\left<X\right>^3$ and thus for a centered random variable with $\left<X\right> = 0$ the first three cumulants are equal to the first three moments. In general, moments and cumulants can be related by the recursion relation
\begin{equation}
C_n(X,t) = \left<X^n\right> - \sum_{m=1}^{n-1} \binom{n-1}{m-1} C_m \left<X^{n-m}\right>\;.
\end{equation}
The numerical computation of the derivatives in Eqs.~\eqref{eq:first_moment_A} and \eqref{eq:cumulants_from_MGF} may lead to various problems, especially at higher orders. Alternatively, the moments and cumulants can be calculated via the derivative-free method introduced in Ref.~\cite{Baiesi2009}.

Although the moments and cumulants provide a rich information about the PDF for $X$, to reconstruct the whole distribution requires knowledge of all the moments and/or cumulants and is thus rarely achievable in practice. For long times $\tau$, however, a very general method for calculating the (approximate) PDF from the MGF can be applied. This method is based on the so-called large deviation theory.
 
\subsection{Large deviation functions}
\label{sec:LDFgeneral}

If the time domain $\tau$ of the time integrals in Eqs.~\eqref{eq:gen_current_variable} and \eqref{eq:gen_noncurrent_variable} gets very large, the PDFs $\rho(X,t_0+\tau,t_0) = \rho(X,\tau)$, $X=A,B$ can assume the so-called large-deviation form \cite{Touchette2009}
\begin{equation}
\log \rho(X,\tau) \sim  \tau J\left(\frac{X}{\tau}\right) \;,
\label{eq:larg_dev_form}
\end{equation}
where the function $J(x)\le 0$ is the large deviation function. The symbol $\sim$ means that Eq.~\eqref{eq:larg_dev_form}
is an asymptotic representation of $\log \rho(X,t_0+\tau,t_0)$ valid for large times $\tau$, where the terms omitted in the formula are typically proportional to $\log \tau$. 

The large deviation function can be calculated from the MGF by Laplace's method. Namely, assuming that $\tau$ is large and Eq.~\eqref{eq:larg_dev_form} holds, the MGF can be written as
\begin{multline}
\log \chi(s_X,\tau) = \log\int dX\, {\rm e}^{-s_X X} \rho(X,\tau)
\approx \\ \log \int dX\, {\rm e}^{-\tau \left[s_X X/\tau - J(X/\tau)\right]}
\approx \tau \max_{x} \left[J(x) - s_X x\right].
\end{multline}
The large deviation function $J(x)$ can hence be calculated by a Legendre--Fenchel transform
\begin{equation}
J(x) = \min_{s_X} \left[ \lambda(s_X) + s_X x
\right]\;, 
\label{eq:LDF_def}
\end{equation}
where 
\begin{equation}
 \lambda(s_X) = \lim_{\tau \to \infty} \frac{1}{\tau}\log \chi(s_X,\tau)
 \label{eq:SCGF}
\end{equation}
denotes the so-called scaled cumulant generating function. Here, we assume that the scaled cumulant generating function is differentiable. Otherwise, the formula~\eqref{eq:LDF_def} does not universally hold, and, one has to resort to a more involved procedure for calculation of the LDF, if it exists at all \cite{Touchette2009}.

For problems with time-independent coefficients and the moment generating function determined by Eq.~\eqref{eq:MGF_TIA} with the tilted Green's function given by $\tilde{{\mathcal U}}(s_X,t_0+\tau,t_0) = \left[\tilde{\mathcal{R}}_{s_X} \tau \right]$, the scaled cumulant generating function~\eqref{eq:MGF_TIA} can be calculated as
\begin{multline}
\frac{1}{\tau}\log \chi(s_X,\tau)
= \frac{1}{\tau}\log\left[ \mathbf{p}^\top_+
\exp\left( \tilde{\mathcal{R}}_{s_X} \tau \right)
\mathbf{p}(t_0)\right] = \\
= \frac{1}{\tau}\log\left\{ \sum_i c_i(s_X) \exp\left[\tau \lambda_i(s_X) \right]\right\}
\approx \lambda_{\rm max}(s_X)\;. 
\label{eq:SCGF_derivati_ti}
\end{multline}
In the calculation, we used the eigenvalue decomposition of the matrix $\tilde{\mathcal{R}}_{s_X}$ which allowed us to rewrite the product $\mathbf{p}^\top_+
\exp\left( \tilde{\mathcal{R}}_{s_X} \tau \right)
\mathbf{p}(t_0)$ using the coefficients $c_i$ arising from products of the vectors $\mathbf{p}^\top_+$, $\mathbf{p}(t_0)$ and eigenvectors of the matrix $\tilde{\mathcal{R}}_{s_X}$. In the final step, we took the limit $\tau \to \infty$ in which the sum is dominated by its largest term $c_{\rm max}\exp(\tau \lambda_{\rm max})$ corresponding the the largest eigenvalue $\lambda_{\rm max}$. In short, the LDF $J(a)$ is in this case determined by the largest eigenvalue $\lambda_{\rm max}(s_X)$ of the tilted rate matrix $\tilde{\mathcal{R}}_{s_X}(s_X)$ as
\begin{equation}
J(x) = \min_{s_X} \left[ \lambda_{\rm max}(s_X) + s_X x
\right]\;.
\label{eq:LDFA}
\end{equation}

For problems with time-dependent coefficients, where the moment generating function is determined by the product form~\eqref{eq:MGF_A} or \eqref{eq:MGF_B}, the large deviation principle~\eqref{eq:larg_dev_form} does not generally hold, unless the time dependence is periodic and we are interested in the PDF for the stochastic variable attained after many cycles $N$ \cite{Barato2018}. For a single cycle starting at $t_0$ and ending at $t_0 + t_c$, where $t_c$ denotes the duration of a single period, the moment generating functions~\eqref{eq:MGF_A} and \eqref{eq:MGF_B} are then determined by the propagator
\begin{equation}
\tilde{\mathcal U}(s) = \lim_{\Delta_t \to 0} \prod_{i=0}^{i_t(t_0+t_c)}
\tilde{{\mathcal U}}_i(s_X)\;.
\label{eq:USA}
\end{equation}
Hence the moment generating function $\chi(s_{X_N})$ for the variable $X_N = X(t_0 + \tau,t_0)$, $\tau=N t_c$ [see Eqs.~\eqref{eq:gen_current_variable} and \eqref{eq:gen_noncurrent_variable}], is given by $\chi(s_{X_N}) = \mathbf{p}^\top_+ \tilde{\mathcal U}(s_{X_N})^N \mathbf{p}(t_0)$. 
A similar calculation as the one in Eq.~\eqref{eq:SCGF_derivati_ti} leads to the scaled cumulant generating function for $X_N$:
\begin{multline}
\frac{1}{\tau}\log \chi(s_{X_N})
= \frac{1}{\tau}\log\left[ \mathbf{p}^\top_+ \tilde{\mathcal U}(s_{X_N})^N \mathbf{p}(t_0) \right] = \\
\frac{1}{\tau}\log\left\{ \sum_i c_i(s_{X_N}) \left[\alpha_i(s_{X_N})\right]^N \right\}
\approx \frac{1}{t_c} \log \alpha_{\rm max}(s_{X_N})\;. 
\label{eq:SCGF_derivati_td}
\end{multline}
Here, $\alpha_i(s_{X_N})$ denote eigenvalues of the propagator for a single cycle $\tilde{\mathcal U}(s_{X_N})$ and the coefficients $c_i$ arise from the products of the vectors $\mathbf{p}^\top_+$, $\mathbf{p}(t_0)$ and eigenvectors of the matrix $\tilde{\mathcal U}(s_{X_N})$. Now, the LDF $J(a)$ is determined by the logarithm of the largest eigenvalue $\alpha_{\rm max}(s_{X_N})$ of the matrix $\tilde{\mathcal U}(s_{X_N})$ as
\begin{equation}
J(x) = \min_{s_{X_N}} \left[\frac{1}{t_c} \log \alpha_{\rm max}(s_{X_N}) + s_{X_N} x
\right]\;.
\label{eq:LDF_td}
\end{equation}
In the fully solved example given in Sec.~\ref{sec:Exmlp} we compute this function for a simple model stochastic heat engine.

\section{Discretization and efficiency}
\label{sec:performance_of_MNM}

There are several ways how to determine suitable discretization
parameters $N_x$, $N_y$ and $N_t$ and the boundaries
$x_{\pm}$ and $y_{\pm}$ without knowing the exact solution. In general, if not fixed by the physics of the problem in question, these parameters should be chosen in such a way that their further refining affects the computed results only negligibly. A second way of choosing the discretization mesh, pursued in the example below, is to compare the numerical results with results obtained using Brownian dynamics (BD) simulations of the stochastic process described by the FPE (\ref{eq:FPGeneral}). Then the mesh can be refined until both methods give the same results. 

For a given discretization, the efficiency (defined as precision of calculation over the computation time) of the MNM is comparable to standard numerical methods based on substituting finite differences for partial derivatives in the FPE \eqref{eq:FPGeneral} such as the one described in Ref.~\cite{Chang1970}. It can be increased by adapting the discretization mesh to the salient features of the time-dependent driving, i.e. by putting the time-discretization parameter $\Delta_t$ roughly inversely proportional to the first derivative of the driving (with some fixed upper bound) and similarly for $\Delta_x$ and $\Delta_y$. 

Main merits of the MNM are: 1) Versatility -- similar implementations can be used for calculating probability distributions, moment generating functions and large deviation functions, both for time-independent and time-dependent problems. 2) Easy implementation -- it is enough to construct the transition rate matrix using the expressions \eqref{eq:ratexEQ}--\eqref{eq:rateyEQ} or \eqref{eq:ratexNEQ}--\eqref{eq:rateyNEQ} and the rest can be handled using matrix operations which are usually well implemented in nowadays programming languages used in physics. And 3) Thermodynamic Consistency -- qualitatively reasonable
predictions of the system dynamics and thermodynamics are obtained with very coarse meshes, as soon as these meshes capture all qualitative features of the forces/potentials and their time dependence. These coarse meshes can thus be used to find interesting effects for a given problem quickly, and thus to reserve time-consuming precise computations for the fraction of model parameters giving the most interesting results. As an example, we refer to the Ref.~\cite{hol17} where all key effects occurring in a complex model of a two-dimensional continuous system were captured by a simple discrete six-level system.

\begin{figure}[t!]
  \centering \includegraphics[width=1.0\linewidth]{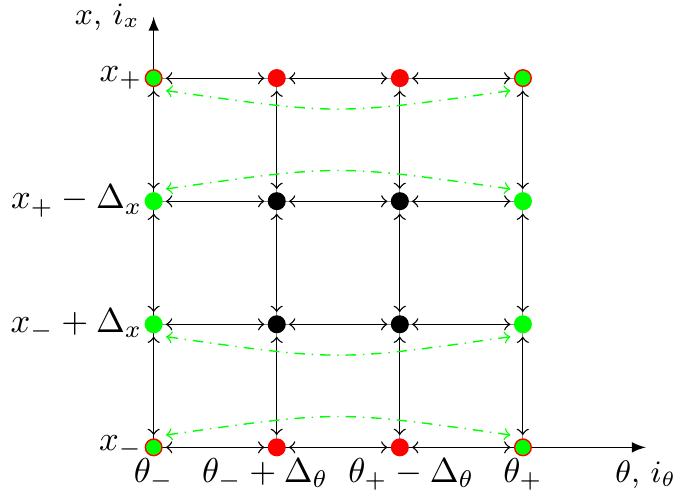}
  \caption{Sketch of the phase space discretization used for the
    numerical solution of the two-dimensional overdamped Fokker-Planck
    equation~(\ref{eq:FPGeneral}) in case of the driven active particle (Sec.~\ref{sec:Exmlp}). The meaning of the arrows and point colors is the same as in Fig.~\ref{fig:space_discretization0}.}
  \label{fig:space_discretization}
  
\end{figure}

The main limitation of the method concerns its generalization to higher-dimensional problems. Namely, the available RAM determines the largest matrix that can swiftly be handled by the computer. The rate matrix $\mathcal R$ in Eq.~(\ref{eq:MasterMatr}) has at most $\prod_{i=1}^d(N_i+1)(1+2d)$ nonzero elements, where $d$
denotes the dimensionality of the problem and $N_i + 1$ denotes the
number of discrete points considered for the $i$th dimension. This is
because each site in Fig.~\ref{fig:space_discretization} is connected
to at most $2d$ neighbors and each of the $\prod_{i=1}^d(N_i+1)$ rows
of $\mathcal R$ thus contains $2d$ rates for transitions into the
given site and $1$ outward rate. On the other hand, the propagators
${\mathcal U}(t,t_0)$ (\ref{eq:Propagator}) already contain
$\prod_{i=1}^d(N_i+1)^2$ nonzero matrix elements. The largest number
of nonzero elements which can be handled by our computer ($8$ GB RAM)
is approximately $10^6$. In practice, problems that can be solved
solely using the rate matrix $\mathcal R$, such as the computation of a (non-equilibrium) stationary solution of Eq.~(\ref{eq:MasterMatr}),
can usually be attacked with acceptable precision in higher dimensions, whereas fully time-dependent problems require additional resources.

\begin{figure}[t!]
  \centering \includegraphics[width=1.0\linewidth]{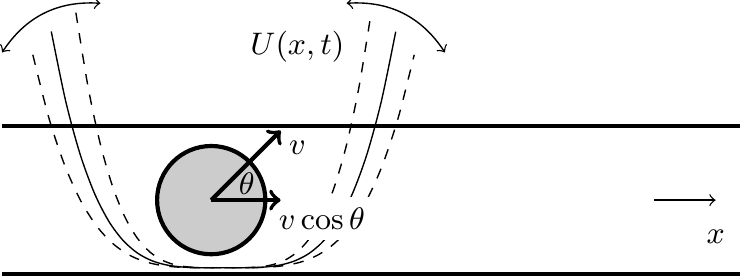}
  \caption{Active particle confined to a single dimension and driven
    by the quartic potential~(\ref{eq:quarticU}) of time-dependent
    strength $k(t)$.}
    \label{fig:example}
\end{figure}

\section{Example: driven active particle}
\label{sec:Exmlp}

An example of a typical application of the MNM can be found in Refs.~\cite{rya16,hol17}, investigating a two-dimensional Brownian ratchet in contact with two reservoirs at different constant temperatures. In this case, the authors used periodic and reflecting boundary conditions. Another example of usage of the MNM is the work~\cite{Ornigotti2017}, where the MNM was used to calculate probability distributions of a particle surviving in a constant unstable cubic potential. In this case, the authors implemented absorbing and reflecting boundary conditions.

In the present section, we consider a FPE with time-dependent coefficients and show that the MNM can be used both for describing the dynamics of the probability distribution and for evaluating MGFs and LDFs of stochastic functionals of the underlying stochastic process. For the sake of simplicity, all physical quantities in this section are represented in suitable natural units that render them dimensionless.

\begin{figure}[t!]
  \centering \includegraphics[width=1.0\linewidth]{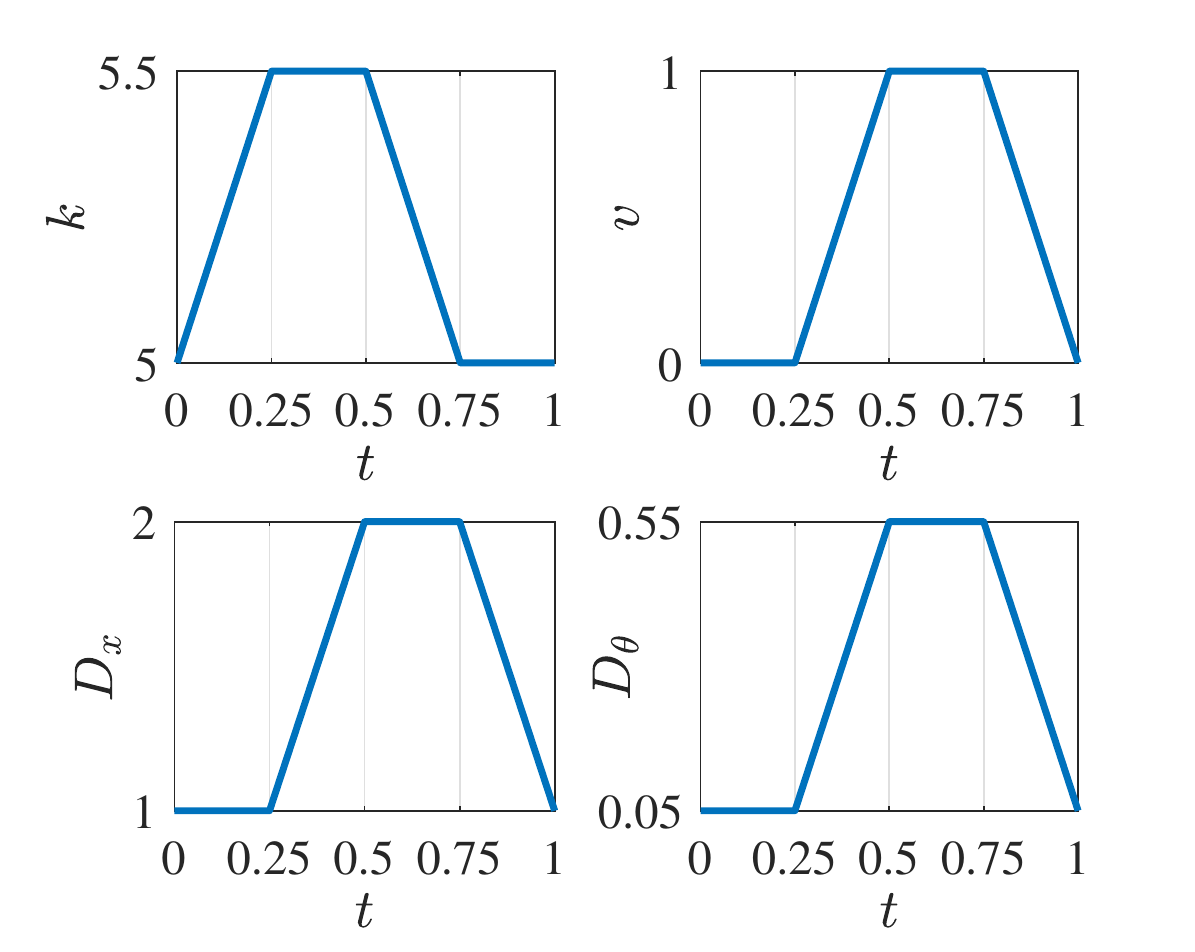}
    \caption{The parameters of the microscopic heat engine consisting of the periodically driven active particle depicted in Fig.~\ref{fig:example}, during one period of the cyclic driving protocol. Boxes comprise maximum and minimum values of the corresponding variables during the cycle.}
      \label{fig:driving}
\end{figure}

We consider an active particle self-propelling with a velocity of
magnitude $v(t) \cos \theta (t)$ and driven by a time-dependent quartic potential
\begin{equation}
U(x,t) = k(t)x^4/4
\label{eq:quarticU}
\end{equation}
in the $x-$direction, as shown in Fig.~\ref{fig:example}. We assume
that the particle motion is overdamped and thus its position $x(t)$
and orientation $\theta(t)$ obey the first-order Langevin equations
\begin{eqnarray}
\dot{x} &=& - k x^3 + v \cos \theta + \sqrt{2D_x} \eta_x\,,
\label{eq:Eq_x}\\
\dot{\theta} &=& \sqrt{2D_\theta} \eta_\theta\,.
\label{eq:Eq_theta}
\end{eqnarray}
Here, $\eta_x$ and $\eta_\theta$ denote independent, zero-mean
unit-variance Gaussian white noises. If we denote the angular variable $\theta$ as $y$, the system \eqref{eq:Eq_x}--\eqref{eq:Eq_theta} corresponds to the FPEs~(\ref{eq:FPGeneral}) and \eqref{eq:FPGeneralTRDB} with $\mu_x F_x = - k x^3 + v \cos \theta$, $\mu_y F_y = 0$, $D_x$ and $D_y = D_\theta$, i.e. 
\begin{equation}
\partial_t \rho = \left[D_x \partial_{x}^2 + D_\theta \partial_{\theta,\theta} - k \partial_{x} x^3  + v \cos \theta \partial_{x} \right] \rho\;,
\label{eq:FPE_HE}
\end{equation}
where $\rho = \rho(x,\theta,t)$. Such schematic models of active particles are often considered as
idealized caricatures of artificial or biological micro-swimmers
\cite{Romanczuk2012,Bechinger2016,epjst:2016}. In fact, they have
acquired the status of a major new paradigmatic toy model of
non-equilibrium statistical mechanics.  While currently most studies
resort to simulations when analytical approximations cease to work
\cite{shi18}, the MNM could in the future provide a welcome
alternative approach.  To illustrate its application to the above
model, we consider a specific non-equilibrium situation that is of
interest for its own sake. Namely, motivated by recent studies
interpreting trapped Brownian particles as microscopic heat engines
\cite{sei12,Holubec2014,Bechinger.2012,Martinez2016,kri16,Verley2014,Polettini2015,Proesmans2015a}, we choose
the potential stiffness $k(t)$, the particle active velocity $v(t)$,
and the diffusion coefficients $D_x(t)$ and $D_\theta(t)$ to be
1-periodic functions, as depicted in Fig.~\ref{fig:driving}. This choice of parameters leads to a positive net work produced by the system per period. 

To understand the thermodynamics of the system, it is helpful to first assume that the particle is not active ($v=0$) and can thus be understood as a system coupled only to a single bath with time-dependent temperature $D_x(t)$. During some parts of the cycle, the heat flows into the bath, during others it flows from the bath to the system. The reservoir with a time-dependent temperature thus serves as a heat source during some parts of the cycle and as a heat sink during the rest of the cycle. Alternatively, one can understand this setup in such a way that there are many reservoirs at different temperatures and the system is at each time connected to one of them. In such case, we would have many heat sources and many heat sinks. In both cases, the laws of thermodynamics allow us to transform heat into work and to operate the system as a heat engine. More details for heat engines of this type can be found in Refs.~\cite{sch08,Holubec2014}. If the particle is active, the basic principle of the engine operation is the same as described above, nevertheless there are some significant differences. Most importantly, the source of the disordered energy (called heat) is now not only the heat bath itself, but also the active self-propulsion of the particle. For more details, we refer the interested reader to Refs.~\cite{kri16,zak17}.

\subsection{Dynamics}

To compute the dynamical and statistical properties of the heat engine using the MNM, we consider the discretization depicted in Fig.~\ref{fig:space_discretization} with $\theta_- = 0$, $\theta_+ = 2\pi - \Delta_\theta$, $\Delta_\theta = 2\pi/(N_\theta + 1)$, and $x_+ = - x_-$, $\Delta_x = (x_+ - x_-)/N_x$. The positions $x_{\pm}$ of the $x$-boundaries of the discrete mesh, where we impose reflecting boundary conditions, are chosen in such a way that the probabilities at the boundaries turn out to be negligible. The discretization parameters $N_\theta$ and $N_x$ are chosen in such a way that their further refinement would barely affect the solution. 

On the discrete lattice, we determine the matrix
${\mathcal U}(t,0)$ in Eq.~(\ref{eq:GreenApprox}), which represents the approximate Green's function for the FPE~\eqref{eq:FPE_HE} of the model, during one driving cycle.
For an arbitrary initial condition ${\mathbf p}_0 = {\mathbf p}(0)$ at time 0, the matrix ${\mathcal U}(t,0)$ provides us with the distribution at time $t$ as
\begin{equation}
{\mathbf p}(t) = {\mathcal U}(t - N,0) \left[{\mathcal U}(1,0)\right]^N {\mathbf p}_0\;,
\end{equation}
where $N = \lfloor t \rfloor$ is the number of full cycles done during the time interval $(0,t)$. After a transient relaxation period, the distribution ${\mathbf p}(t)$ becomes independent of the initial condition. As a consequence of the periodicity of the driving, it converges to a 1-periodic vector in the long-time limit.

\begin{figure}[t!]
\centering \includegraphics[width=1.0\linewidth]{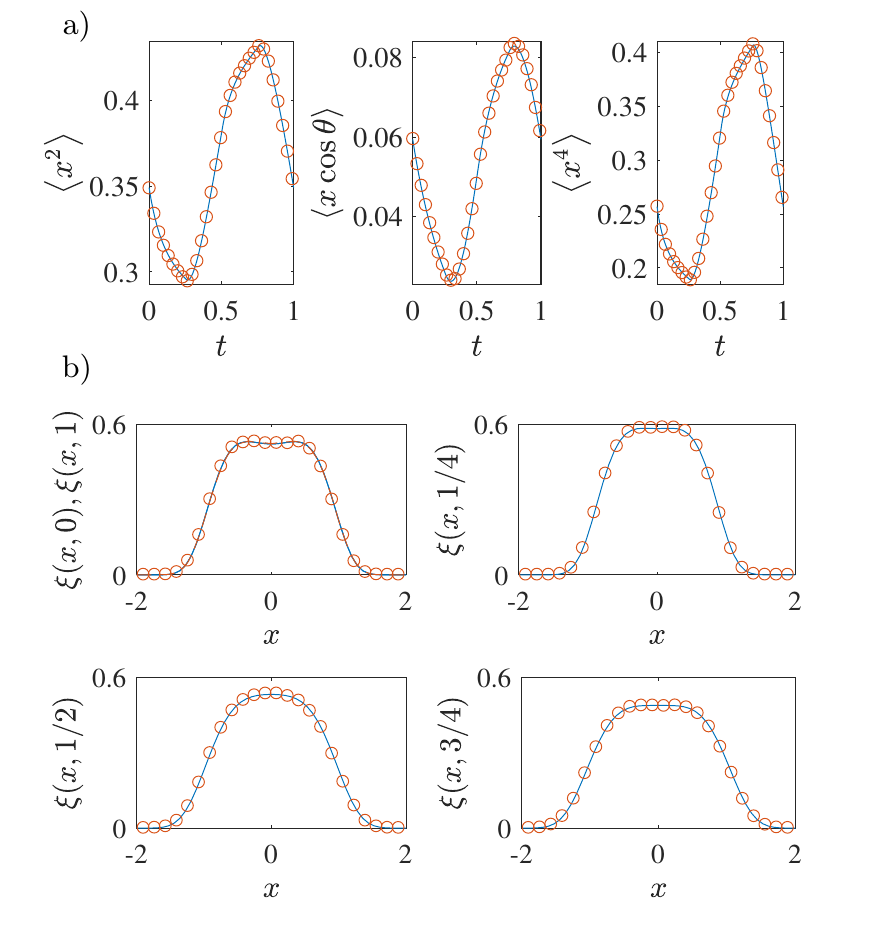}
\caption{Comparison of observables for the periodically driven active
  particle, depicted as a function of time during one limit cycle, as
  computed from BD simulations (symbols) and the
MNM (lines): a) averages $\left<x^2\right>$, $\left<x \cos
  \theta\right>$ and $\left<x^4\right>$; b) marginal probability
  density $\xi(x,t)$ for the particle position $x$ at five time-instants $t$ during the
  cycle. The PDF $\xi(x,0)$ at the initial time $0$ (dashed blue line) and $\xi(x,1)$ at the final time $1$ (full orange line) in the first panel of b) coincide, because the system operates in the limit cycle as described by Eq~\eqref{eq:limit_cycle}.}
\label{fig:dynamics}
\end{figure}

This time-dependent long-time solution ${\mathbf p}_{\rm lc}(t)$ of the Master equation (FPE) with periodic transition rates is called the limit cycle. Using its periodicity, it can be determined using the eigenvector of the Green's function ${\mathcal U}(t,0)$ corresponding to the eigenvalue 1 as
${\mathbf p}_{\rm lc}(1)={\mathcal U}(1,0) {\mathbf p}_{\rm lc}(0) = {\mathbf p}_{\rm lc}(0)$,
\begin{equation}
{\mathbf p}_{\rm lc}(t) = {\mathcal U}(t,0) {\mathbf p}_{\rm lc}(0)\;.
\label{eq:limit_cycle}
\end{equation}
From this approximate solution and the relation
(\ref{eq:mapping_rho_p}), we compute the approximate probability distribution $\rho(x,\theta,t)$ of the active particle during the engine's operation. We use it to numerically compute the averages $\left<x^2\right>$, $\left<x \cos
\theta\right>$ and $\left<x^4\right>$ as functions of time and the
marginal distribution for the $x$-position $\xi(x,t) = \int_0^{2\pi} d\theta
\rho(x,\theta,t)$ at five time instants $t=0,1/4,1/2,3/4$ and $1$,
during the limit cycle. We also independently evaluated these
quantities using a BD simulation of the system \eqref{eq:Eq_x}--\eqref{eq:Eq_theta}. The comparisons of the
averages and the marginal distributions are shown in
Figs.~\ref{fig:dynamics}a/b, respectively. The MNM results, depicted
by full lines, perfectly overlap with those of the BD (symbols).
The MNM results were calculated using the discretization parameters $N_x = 51$, $N_\theta = 21$, $N_t = 76$ and $x_\infty = 2.4$. Already for $N_x = 31$, $N_\theta
= 15$, $N_t = 76$ and $x_\infty = 2.4$ one obtains curves that are
visually indistinguishable from those depicted in
Fig.~\ref{fig:dynamics}, while the calculation is approximately 10 $\times$ faster than with the finer mesh. For the BD we generated $10^6$ trajectories with the integration step $10^{-3}$. 

Besides checking the correctness of our implementation of the MNM by BD, we have also tested our numerical results against analytical results available for the presented model in two limiting situations. Specifically, we tested that the computed PDF attains the form $\rho(x,\theta,t) \propto \exp[-U(x,t)/T_{\rm eff}(t)]$, $T_{\rm eff} = T + v^2/(2 D_\theta)$ for a quasi-static driving and $D_\theta \gg 1$. In this case, the particle rotates so fast that the term $v \cos \theta$ in Eq.~\eqref{eq:Eq_x} becomes equivalent to a further white noise with the effective temperature $v^2/(2 D_\theta)$. As a second benchmark, we considered quasi-static driving with $D_\theta \to 0$, where the active velocity can be treated as constant and thus $\rho(x,\theta,t) \propto \exp\left\{[-\left[U(x,t) - v x \cos \theta\right]/T\right\}$.

\subsection{Moment generating functions}
\label{sec:MGF}

Besides computing the distribution $\rho(x,\theta,t)$ to evaluate averages, moments, and reduced distribution functions for $x$ and $\theta$, the MNM can also be applied directly to other comprehensive representations of the stochastic thermodynamics encoded in the FPE. In the following, we apply the MNM to directly compute
moment generating functions (MGFs) and large-deviation functions
(LDFs) of work and heat. From the point of view of stochastic thermodynamics, these MGFs and LDFs are of interest in studies of
work fluctuations in microscopic heat engines operating close to the reversible efficiency \cite{Holubec2018a,Pietzonka2018,Koyuk2018} or of the fluctuating efficiency \cite{Verley2014,Polettini2015,Proesmans2015a}, both intensely investigated during the last few years. 

In stochastic thermodynamics of externally driven systems, work and heat are usually defined from the first law of thermodynamics, as follows \cite{sek10,sei12}. The energy $U(x,t)$ of the particle in a fixed micro-state $(x,\theta)$ can change in the course of time in two fundamentally different ways, one called work $w$, the other heat $q$. Formally, we can write $dU(x,t)/dt = \dot{w}(x,t) + \dot{q}(x,t)$, where
  \begin{eqnarray}
  \dot{w}[x(t),t] = \dot{w}(t) &\equiv & \dot{k}(t) x^4(t)/4\;,
\label{eq:Eq_dw}\\
\dot{q}[x(t),t] = \dot{q}(t) &\equiv & k(t)x^3(t)\dot{x}(t)\;.
\label{eq:Eq_dq}
\end{eqnarray}
The work done on the particle per unit time, $\dot{w}$, is thus nonzero only if the potential is externally changed [$\dot{k}(t) \neq 0$]. A heat exchange $|\dot{q}|>0$ occurs if the particle moves in the potential and either dissipates its kinetic energy or transforms energy acquired from the bath or from the active self-propulsion into potential energy. Since the considered particle is active, there is necessarily also some dissipated energy [mostly much larger than \eqref{eq:Eq_dq}] related to the self-propulsion mechanism. This energy is usually called housekeeping heat, and we neglect it here, treating it as an intrinsic property of the system.

Work and heat flowing to the particle during the time interval $(0,\tau)$ are defined as integrals over the respective rates \eqref{eq:Eq_dw} and \eqref{eq:Eq_dq}:
\begin{eqnarray}
  w(\tau) &=& \int_0^\tau \!\!\text{d}t \,\dot{w}(t) = \int_0^\tau \!\!\text{d}t \,\partial_t U[x(t),t] \;,
\label{eq:Eq_w}\\
q(\tau) &=& \int_0^\tau \!\!\text{d}t \, \nabla U[x(t),t]\cdot [x(t),\theta(t)]
\nonumber \\
&= & [U(x(\tau),\tau) - U(x(0),0)] - w(\tau)\;.
\label{eq:Eq_q}
\end{eqnarray}
They correspond to the cumulative external work
performed on the active particle by the device varying the confinement strength, and the cumulative heat transferred to it from the thermal reservoir at the time-dependent temperature $T$. Additionally, the energy gained due to the self-propulsion of the swimmer is counted as (``internal'' or ``active'') heat supply. The cumulative work is an example of a variable that is not proportional to the probability current, with the function $b[x(t),\theta(t),t]$ in Eq.~\eqref{eq:gen_noncurrent_variable} given by the instantaneous potential energy of the particle multiplied by the total time $\tau$, $b = \tau U[x(t),t]$. The cumulative heat, one the other hand, is an example of an observable proportional to the current, with the vector $\mathbf{c}(x,y,t)$ in Eq.~\eqref{eq:gen_current_variable} given by $\tau \nabla U[x(t),t]$.

\begin{figure}[t!]
\centering \includegraphics[width=1.0\linewidth]{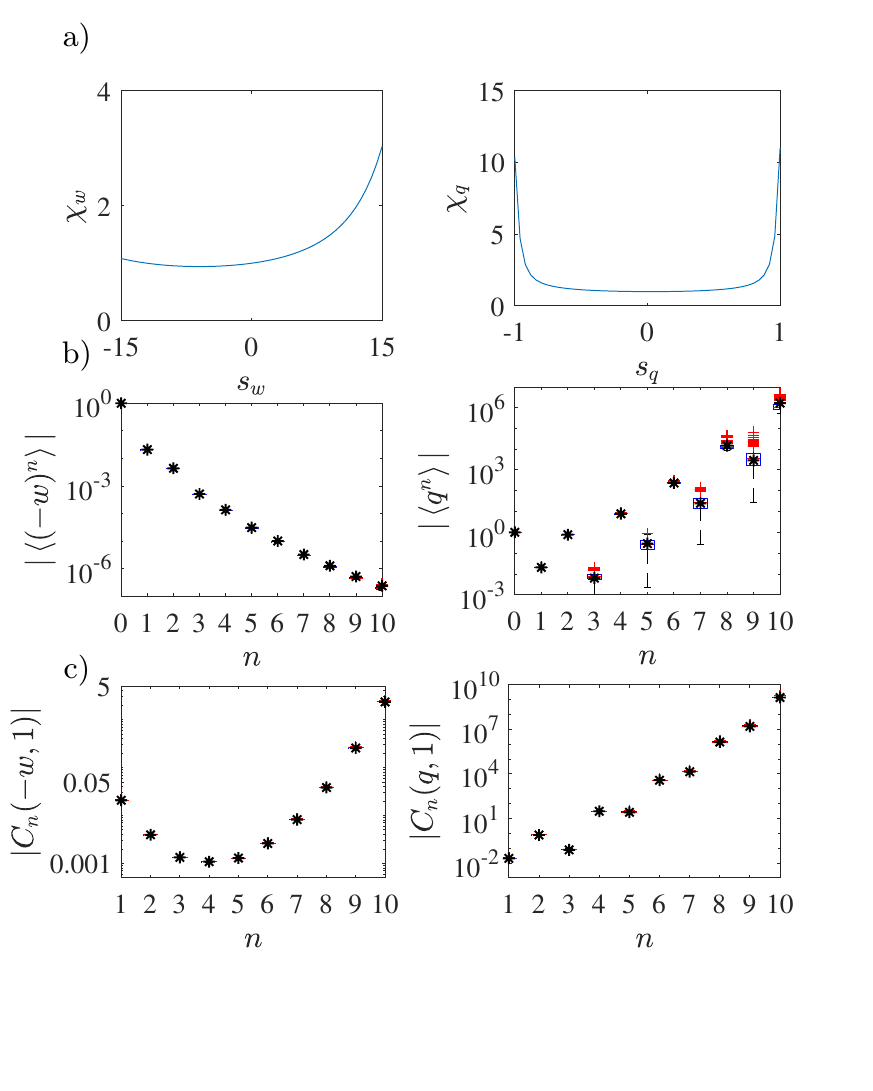}
\caption{Application of the MNM to moment generating functions (MGFs):
  a) the MGFs $\chi_w$ and $\chi_q$ for work and heat; b) the first 11
  raw moments and c) the first 10 cumulants [c)] of the net extracted work $-w$ and net supplied heat $q$
  per cycle, as calculated from the MGFs depicted in panel a)
  ({\Large$\ast$}) and from corresponding BD simulations of $200\times 10^6$
  trajectories (box plots).}
\label{fig:moments}
\end{figure}

\begin{figure}[t!]
\centering \includegraphics[width=1.0\linewidth]{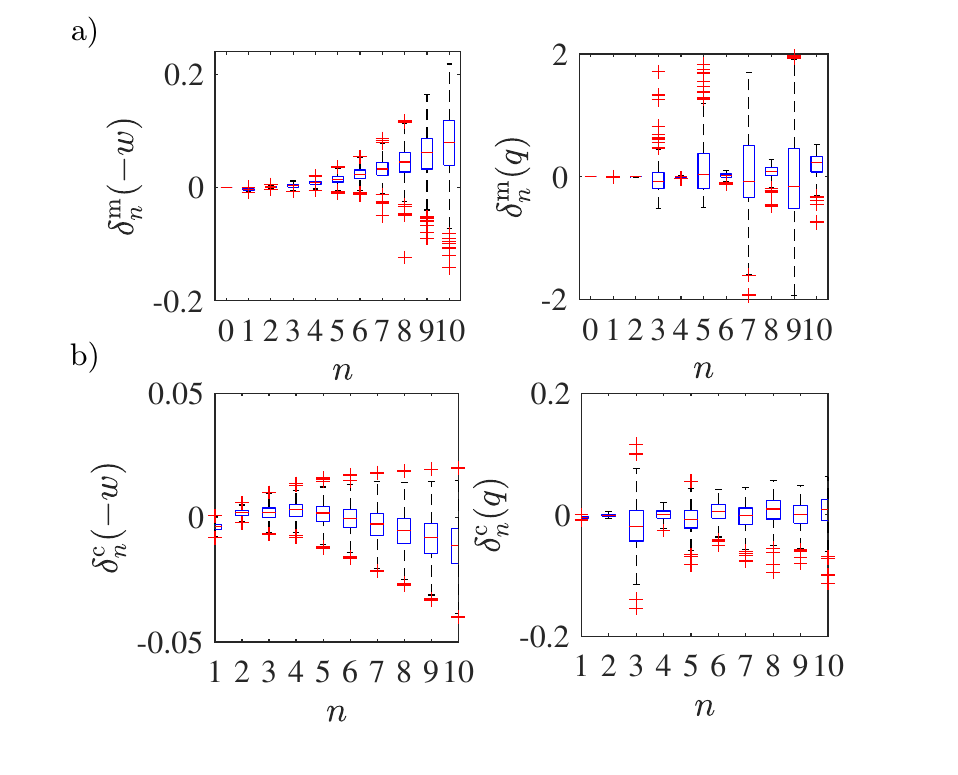}
\caption{Box plots of relative differences between moments \eqref{eq:delta_moments} [panels a)] and cumulants \eqref{eq:delta_cumulants} [panels b)] for work (left) and heat (right) as computed using MNM and BD, respectively (see Fig.~\ref{fig:moments}b, c).}
\label{fig:moments_errors}
\end{figure}

Consider now the driving protocol depicted in Fig.~\ref{fig:driving} and the discretized time according to Sec.~\ref{sec:time_dependent}.
Using the formula \eqref{eq:MGF_B} of Sec.~\ref{sec:MGF}, we calculated the MGF $\chi_w = \chi_w(s_w)$ for the work $w(1)$ [Eq. (\ref{eq:Eq_w})] transferred to the active particle during one limit cycle. The corresponding MGF $\chi_q = \chi_q(s_q)$ for the heat $q(1)$ [Eq.~(\ref{eq:Eq_q})] follows from formula \eqref{eq:MGF_A} with tilted matrices $\tilde{{\mathcal U}}_i(s_q) = \exp\left[ \tilde{\mathcal{R}}_{s_q}(t_0 + i \Delta_t) \Delta_t \right]$ if $i \ge 1$ and $\tilde{{\mathcal U}}_0 = {\mathcal I}$ otherwise. For the parts of the piece-wise constant protocol with time-independent potential, the tilted matrices can also be computed from the formula~\eqref{eq:UAti}.

The resulting moment generating functions are shown in Fig.~\ref{fig:moments}a. The MGFs were sampled for $s_w \in (-15, 15)$ with the step $\Delta_{s_w} = 3/5$ for work and for $s_q \in (-1, 1)$ with the step $\Delta_{s_q} = 2/50$ for heat. To check the results, we computed the first 11 raw moments using the formula~\eqref{eq:moments_from_MGF} and the first 10 cumulants using the formula~\eqref{eq:cumulants_from_MGF}. For the numerical evaluation of the derivatives in these equations we used the central difference scheme
\begin{equation}
\frac{d^n f(z)}{dz} \approx \sum_{i=0}^n (-1)^i \binom{n}{i}
f\left[z + \left(\frac{n}{2} - i \right)\Delta_z\right]\;,
\end{equation}
where $f$ is given by $\chi_w$ for moments/cumulants of work and by $\chi_q$ for moments/cumulants of heat. The parameters $z$ and $\Delta_z$ are given by $s_w$ and $2 \Delta_{s_w}$ for $\chi_w$ and $s_q$ and $2\Delta_{s_q}$ for $\chi_q$.

The resulting moments are depicted in Fig.~\ref{fig:moments}b ({\Large $\ast$}) together with the
corresponding results obtained from the BD simulations (depicted using box plots~\footnote{On each blue box, the central red mark indicates the median, and the bottom and top blue edges of the box indicate the 25th and 75th percentiles, respectively. The black dashed whiskers extend to the most extreme data points not considered outliers, and the outliers are plotted individually using the red '+' symbol. Taken from Matlab documentation.}). In order to assess the error of the latter, we simulated
each moment 200 times using $10^6$ trajectories yielding a box plot for each $n$ in the figure. For the exchanged work, all data {\Large $\ast$} from the MNM and the corresponding box plots from BD perfectly superimpose so that the box plots are hardly visible, for all values of $n$ (Fig.~\ref{fig:moments}b, left). For heat, the results from both methods either coincide, or the MNM results lie within the boxes indicating the 25th and 75th percentiles of the BD data (Fig.~\ref{fig:moments}b, right). The cumulants resulting from the MNM depicted in Fig.~\ref{fig:moments}c ({\Large $\ast$}) together with the corresponding results obtained from the BD simulations (box plots) agree both for work (Fig.~\ref{fig:moments}c, left) and for heat (Fig.~\ref{fig:moments}c, right). Note that the computation of cumulants from BD simulation is much less demanding than the computation of moments due to suppressed fluctuations.

To get a better insight into the precision of theses results, we show in Fig.~\ref{fig:moments_errors} box plots of relative differences
\begin{equation}
\delta^{\rm m}_n(x) = \frac{\left<x^n\right>_{\rm a} - \left<x^n\right>_{\rm s}}{\left<x^n\right>_{\rm a} + \left<x^n\right>_{\rm s}}
\label{eq:delta_moments}
\end{equation}
of computed and simulated moments for work ($x = -w$, Fig.~\ref{fig:moments_errors}a, left) and heat ($x = q$, Fig.~\ref{fig:moments_errors}a, right) and relative differences
\begin{equation}
\delta^{\rm c}_n(x) = \frac{C^{\rm a}_n(x,1) - C^{\rm s}_n(x,1)}{C^{\rm a}_n(x,1) + C^{\rm s}_n(x,1)}\;,
\label{eq:delta_cumulants}
\end{equation}
of computed and simulated cumulants of work ($x = -w$, Fig.~\ref{fig:moments_errors}b, left) and heat ($x = q$, Fig.~\ref{fig:moments_errors}b, right). The relative differences for work increase with $n$ showing a trend towards positive relative differences for moments and negative relative differences for cumulants. These trends are caused by the chosen discretization. For heat, the data from BD are much more noisy than those for work and therefore no trend in the relative differences is detectable. Even with the obvious trends in the relative differences for work, all the data shown in Fig.~\ref{fig:moments_errors} are relatively well centered around 0 showing a good agreement between the results computed using the MNM and the BD.

\subsection{Large--deviation functions}
\label{sec:LDF}

Let us now investigate fluctuations of work $w(\tau N)=w=\int_0^{\tau N} \!\!\text{d}t \,\dot{w}(t)$ and heat $q(\tau N)=q=\int_0^{\tau N} \!\!\text{d}t \,\dot{q}(t)$ integrated over many cycles $N\gg 1$ of duration $\tau = 1$ [see Eqs.~\eqref{eq:Eq_w}--\eqref{eq:Eq_q}]. According to the large deviation theory \cite{Touchette2009} reviewed in Sec.~\ref{sec:LDFgeneral}, in such situation the PDFs for work and heat assume the form \eqref{eq:larg_dev_form} with $X/\tau = w/\tau N$ and $X/\tau = q/\tau N$ for work and heat, respectively, on the right-hand side, i.e.
\begin{eqnarray}
\rho_w(w) & \sim & \exp\left[\tau N J_w\left(\frac{w}{\tau N}\right)\right]\;,
\label{eq:rhowHELDF}\\
\rho_q(q) & \sim & \exp\left[\tau N J_q\left(\frac{q}{\tau N}\right)\right]\;.
\label{eq:rhoqHELDF}
\end{eqnarray}
The LDFs $J_w(w)$ and $J_q(q)$ are determined by the largest eigenvalues of the tilted propagators used in the previous section for the MGFs, see Sec.~\ref{sec:LDFgeneral} and Eq.~\eqref{eq:USA}--\eqref{eq:LDF_td} for details.

\begin{figure}[t!]
\centering \includegraphics[width=1.0\linewidth]{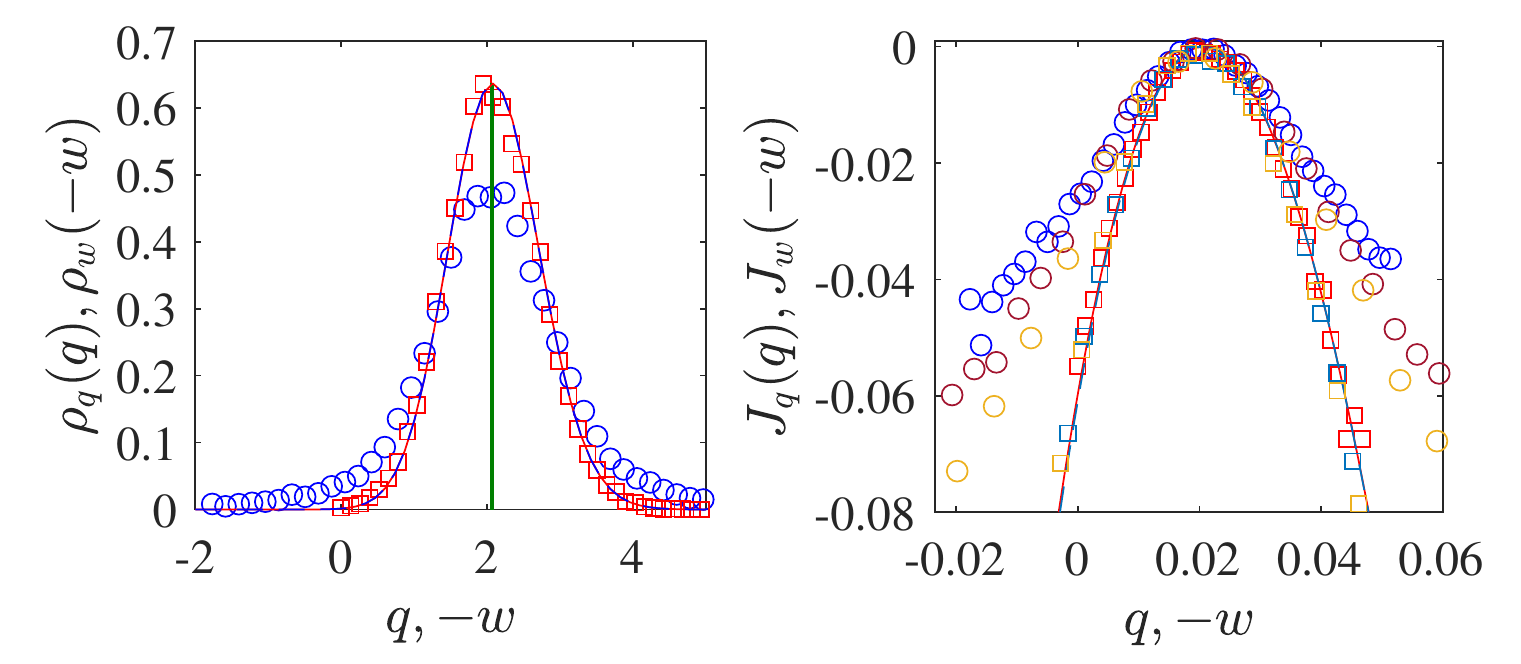}    \caption{Large-deviation limit of heat and work distributions.  BD simulations of the cumulative distributions $\rho_q$ and $\rho_w$ of the net heat $q$ supplied ($\bigcirc$) and the net work $-w$ extracted ($\square$) over 100 cycles show that the work distribution has converged to the common limiting form obtained from the MNM (superimposing lines, with the vertical line indicating the average), while the heat distribution has not (left panel). The Legendre--Fenchel transformed logarithms of the MGF of heat and work (right panel) elucidate the unequal convergence towards the common large-deviation function (LDF) $J_q(q) \sim J_w(-w)$ with the number of cycles $N=30$, $50$, $100$. While the work distribution ($\square$) has already converged for $N=30$, the heat distribution ($\bigcirc$) keeps evolving (top to bottom).}
  \label{fig:LDF}
\end{figure}

In Fig.~\ref{fig:LDF}, we show the LDFs $J_w(w)$ and $J_q(q)$ computed using the MNM. For $N\gg 1$, the nonextensive boundary term
$U[x(t),t] - U[x(0),0]$ in Eq.~(\ref{eq:Eq_q})
can be neglected as compared to $-w(\tau N)$, so that $\rho_x(x) \sim \exp[\tau N J_x(x/\tau N)]$ for $x=q,-w$, and $J_q(q) = J_w(-w)$, as is verified by our MNM results
(superimposing lines). However, the data obtained from $10^6$ BD
trajectories (symbols) shows that only the work distribution
($\square$) attains the large deviation limit quickly, while the heat
distribution ($\bigcirc$) has not converged, even for $N=100$ cycles. This is because, for the parameters considered in our numerical study, heat fluctuates much more than work, as already suggested by the moments and cumulants shown in Fig.~\ref{fig:moments}b and c. Let us note that while we have computed the LDFs using the standard BD, which was much more time consuming than the evaluation of the MNM, there are various optimized simulation algorithms~\cite{Giardina2006,Nemoto2016,Ferre2018} for computing of LDFs that can render BD simulations more competitive.

\section{Conclusion and outlook}
\label{sec:Concl}

We have presented a numerical scheme for overdamped FPEs with time-dependent coefficients, based on the mapping between the FPE and a Master equation with detailed-balanced transition rates. The resulting numerical method yields thermodynamically consistent results for arbitrary discretizations. It can be used for solving the FPE and also for computing MGFs and LDFs for functionals defined along the trajectories of the stochastic process underlying the FPE. 

The performance of the method for solving the FPE is similar to other numerical methods relying on approximating the derivatives by finite differences. However, due to its thermodynamic robustness, the method predicts well the qualitative behavior of the studied system already for coarse meshes that capture merely the salient features of the force field/potential landscape. Thus the MNM can safely be used for a fast scanning of the parameter space if one looks for interesting effects. 

The presented numerical scheme shares basic notions with so-called Markov-state
models of molecular kinetics, which have been employed for
interpreting data from single-molecule experiments and
molecular-dynamics simulations \cite{Prinz2011}.  Both methods exploit
the mapping of stochastic processes occurring in continuous space and
time to discrete state-space Markov processes. While the kinetic
Markov-state models are often based on special protocols, such as
time-periodic driving \cite{Wang2015}, our formulation can in
principle handle arbitrary time-dependent protocols. 

Unfortunately, the MNM cannot easily be generalized to underdamped systems because it relies on the mapping~\eqref{eq:mapping_rho_p} between the FPE~\eqref{eq:FPGeneralTRDB} and the Master equation~\eqref{eq:MasterEQ}, which is restricted to overdamped dynamics. The difficulties with the underdamped limit can be anticipated from the transition rates~\eqref{eq:ratexEQ}--\eqref{eq:rateyEQ} and \eqref{eq:ratexNEQ}--\eqref{eq:rateyNEQ}, which are all of the form $D\exp(\pm A/D)$. Thus some of them necessarily diverge if the diffusion coefficient $D$ goes to zero. The only variables with vanishing diffusion coefficient (noise) in the Langevin equation [see e.g. Eqs.~\eqref{eq:Eq_x}--\eqref{eq:Eq_theta} for variables with nonzero diffusion coefficients in the Langevin equation] tractable by the MNM in its present form are variables like work and heat [see Eqs.~\eqref{eq:Eq_dw}--\eqref{eq:Eq_dw}], which do not feed back onto the dynamics of the noisy variables. For such variables, the MNM yields MGFs and LDFs. The presented form of the MNM is thus limited to those underdamped situations where the momentum in the underdamped Langevin equation does not depend on the position. A promising way to generalize the MNM for general underdamped dynamics could build on the path integral method suggested in Ref.~\cite{Lyasoff2003}, which shares with the MNM the important property of summing over all possible paths of the stochastic process and thus allows to naturally incorporate calculations of various path-dependent stochastic variables. Another possible pathway to generalize the MNM to underdamped systems may be to incorporate into the MNM the ideas used in the formulation of the lattice Boltzmann method \cite{He1997,Chen1998}.

\appendix
\section{Space dependent diffusion coefficient}
\label{app:DFC}

In this appendix, we will show that the transition rates obeying the local detailed balance condition \eqref{eq:rates0} can be used for solving only FPEs with position independent diffusion coefficients. For the proof, it suffices to consider the one dimensional FPE
\begin{equation}
\partial_t \rho(x,t) = \left[\partial_{x}^2 D_x(x,t) - \partial_{x} \mu_x  F_x(x,t) \right] \rho(x,t)\;
\label{eq:FP1D}
\end{equation}
and the corresponding Master equation
\begin{multline}
\dot{p}_{i_x} = r_{i_x+1 \to i_x}p_{i_x+1} + r_{i_x-1 \to i_x}p_{i_x-1} - \\
\left(r_{i_x \to i_x + 1} + r_{i_x \to i_x - 1} \right)p_{i_x}
\label{eq:MasterEQ1D}
\end{multline}
on the discrete lattice with points indexed by $i_x = {\lfloor}\frac{x - x_-}{\Delta_x}{\rfloor}$ and the lattice parameter $\Delta_x = \frac{x_+ - x_-}{N_x}$. We assume that $D_x$ and $F_x$ in \eqref{eq:FP1D} depend on time $t$ and position $x$ and we look for transition rates in \eqref{eq:MasterEQ1D} fulfilling the condition \eqref{eq:xrates} and yielding Eq.~\eqref{eq:FP1D} in the leading order in the discretization parameter $\Delta_x$ if we set $\rho(x,t) = \lim_{\Delta_x \to 0} p_{i_x}/\Delta_x$. 

In one dimension, the entropy production $\Delta S_{\rm R} (x \to x + \Delta_x) = \Delta S_{\rm R} = \int_{x}^{x+\Delta_x}dx'\, \frac{F_x(x')}{T_x(x')}$ along the transition from $x$ to $x + \Delta_x$ can be written as 
\begin{eqnarray}
\Delta S_{\rm R}/k_{\rm B} = -\left[\tilde{U}(x+\Delta x,t)-\tilde{U}(x,t)\right]\;,
\end{eqnarray}
where $\tilde{U}$ is a dimensionless potential such that $F_x/k_{\rm B} T_x = \mu_x F_x/D_x = -\partial_x \tilde{U}$. The transition rates satisfying the detailed balance condition \eqref{eq:rates0} can thus in general be written as
\begin{eqnarray}
r_{i_x \to i_x + 1} &=& \frac{A_{i_x+1/2}}{\Delta_x^2}\exp{\left[-\frac{\tilde{U}_{i_x + 1}-\tilde{U}_{i_x}}{2}\right]}\;,
\label{eq:rateA1}
\\
r_{i_x+1 \to i_x}  &=& \frac{A_{i_x+1/2}}{\Delta_x^2}\exp{\left[\frac{\tilde{U}_{i_x + 1}-\tilde{U}_{i_x}}{2}\right]}\;,
\label{eq:rateA2}
\end{eqnarray}
where $\tilde{U}_{i_x} = \tilde{U}(x_- + \Delta_x i_x,t)$, and $A_{i_x+1/2} = A[x_- + \Delta_x (i_x + 1/2),t]$ is some space-and-time dependent function determining the prefactor of the transition rates. Inserting the rates \eqref{eq:rateA1}--\eqref{eq:rateA2} in the Master equation \eqref{eq:MasterEQ1D} we obtain up to the leading order in $\Delta_x$ a partial differential equation of the form
\begin{equation}
\partial_t{\rho} = A'(\rho \tilde{U}' + \rho') + A(\rho'\tilde{U}' + \rho'' + \rho \tilde{U}'')\;,
\label{eq:MEA1}
\end{equation}
where $\rho' \equiv \partial_x \rho(x,t)$. The first nonzero correction to Eq.~\eqref{eq:MEA1} is of order $\Delta_x^2$. Comparing Eq.~\eqref{eq:MEA1} with the desired Eq.~\eqref{eq:FP1D} and using $\tilde{U}' = -\mu_x F_x/D_x$, we find that it is not possible to choose $A(x)$ in such a way that the two equations are identical, unless the diffusion coefficient is position independent ($D_x' = 0$) and $A = D_x$.

The main problem why the transition rates of the form~\eqref{eq:rateA1}--\eqref{eq:rateA2} can not yield the FPE~\eqref{eq:FP1D} with space dependent coefficients are the prefactors $A_{i_x+1/2}$, which must be the same for the transitions $i_x \to i_x + 1$ and $i_x + 1 \to i_x$. If we relax this assumption [and thus we do not consider only the rates strictly fulfilling the local detailed balance condition \eqref{eq:rates0}], it is not difficult to find transition rates which can be used for solving the FPE~\eqref{eq:FP1D} in its full generality. They read
\begin{eqnarray}
r_{i_x \to i_x + 1} &=& \frac{A_{i_x}}{\Delta_x^2}\exp{\left[-\frac{\tilde{U}_{i_x + 1}-\tilde{U}_{i_x}}{2}\right]}\;,
\label{eq:rateA3}
\\
r_{i_x+1 \to i_x}  &=& \frac{A_{i_x + 1}}{\Delta_x^2}\exp{\left[\frac{\tilde{U}_{i_x + 1}-\tilde{U}_{i_x}}{2}\right]}\;,
\label{eq:rateA4}
\end{eqnarray}
with $\tilde{U}_{i_x} = \tilde{U}(x_- + \Delta_x i_x,t)$ and $A_{i_x} = A(x_- + \Delta_x i_x,t)$, where $\tilde{U}' = -\mu_x F_x/D_x$ and $A(x,t) = D_x(x,t)$. Inserting these transition rates into the Master Eq.~\eqref{eq:MasterEQ1D} we obtain up to the leading order in $\Delta_x$ the FPE~\eqref{eq:FP1D}. The first nonzero correction is of the order of $\Delta_x^2$. Although the rates \eqref{eq:rateA3}--\eqref{eq:rateA4} do not obey the strict local detailed balance condition \eqref{eq:rates0}, they still describe dynamics that conserves positivity and normalization. Furthermore, for position-independent diffusion coefficients, the detailed-balanced rates \eqref{eq:rateA1}--\eqref{eq:rateA2}  and the rates \eqref{eq:rateA3}--\eqref{eq:rateA4} are identical. The generalization of the transition rates \eqref{eq:rateA3}--\eqref{eq:rateA4} to higher dimensions is straightforward.

\section{Moment generating function for time-averaged current}
\label{app:MGF_intergated _current}

In this appendix, we calculate the moment generating function $\chi_{\bar{\mathfrak{J}}(\mathbf{r}_a)}(\mathbf{s}_{\bar{\mathfrak{J}}},\tau) = \chi_{\bar{\mathfrak{J}}(\mathbf{r}_a)}(s_{\bar{\mathfrak{J}}_x},s_{\bar{\mathfrak{J}}_y},\tau)$ for the time-averaged particle current trough the position $\mathbf{r}_a = (x_a,y_a)$ at time $t_0$:
\begin{equation}
\bar{\mathfrak{J}}(\mathbf{r}_a,\tau) = \bar{\mathfrak{J}}(\mathbf{r}_a,\tau,t_0) = \frac{1}{\tau}\int_{t_0}^{t_0+\tau}dt \mathfrak{J}(\mathbf{r}_a,t)\;.
\label{eq:v_average_ra}
\end{equation}
In the limit $\tau \to 0+$ this random variable converges to the microscopic current $\mathfrak{J}(\mathbf{r}_a,t)$ defined in Eq.~\eqref{eq:micro_current}. The following strategy for calculating $\chi_{\bar{\mathfrak{J}}(\mathbf{r}_a)}(\mathbf{s}_{\bar{\mathfrak{J}}},\tau)$ can be easily generalized to more complex random variables discussed in Sec.~\ref{sec:MGF_current}.

The MGF $\chi_{\bar{\mathfrak{J}}(\mathbf{r}_a)}(\mathbf{s}_{\mathfrak{J}},\tau)$ is defined as the two-sided Laplace transform
\begin{equation}
\chi_{\bar{\mathfrak{J}}(\mathbf{r}_a)}(\mathbf{s}_{\bar{\mathfrak{J}}},\tau) 
= \int d\bar{\mathfrak{J}}_x \int d\bar{\mathfrak{J}}_y {\rm e}^{-\mathbf{s}_{\bar{\mathfrak{J}}}\cdot\bar{\mathfrak{J}}}
p_{\bar{\mathfrak{J}}(\mathbf{r}_a)}(\bar{\mathfrak{J}},\tau)
\label{eq:MGF_def}
\end{equation}
of the probability distribution $p_{\bar{\mathfrak{J}}(\mathbf{r}_a)}(\bar{\mathfrak{J}},\tau)$ for $\bar{\mathfrak{J}}(\mathbf{r}_a)$. To calculate an approximation to $\chi_{\bar{\mathfrak{J}}(\mathbf{r}_a)}(\mathbf{s}_{\bar{\mathfrak{J}}},\tau) $ using the discrete model of Fig.~\ref{fig:space_discretization0}, we count the number $n_x^r$ of jumps to the right from the site $(i_{x_a},i_{y_a})$ during the time interval $(t_0,t_0+\tau)$ and also the corresponding number $n_x^l$ of jumps to the left from the site $(i_{x_a} + 1,i_{y_a})$ to get the net transport
\begin{equation}
n_x = n_x(\tau,t_0) = n_x^r - n_x^l = \lim_{\Delta_x \to 0, \Delta_y \to 0} \left( \tau \Delta_y \bar{\mathfrak{J}}_x \right)\;.
\label{eq:vx_aprr}
\end{equation}
Here, the factor $\Delta_y$ comes from Eq.~\eqref{eq:current_ME1}. Similarly, the numbers $n_y^u$ and $n_y^d$ of jumps up from $(i_{x_a},i_{y_a})$ and down from $(i_{x_a},i_{y_a} + 1)$, respectively, determine $n_y = n_y^u - n_y^d = \lim_{\Delta_x \to 0, \Delta_y \to 0}\left( \tau \Delta_x \bar{\mathfrak{J}}_y \right)$. 

Let us now consider a time interval $dt$ so short that only a single jump can occur and investigate the PDF for $\mathbf{n}=(n_x,n_y)$, which can be mapped to the PDF for the time-averaged current $\bar{\mathfrak{J}}$. At the initial time $t_0$, the distribution of the particle position is described by the vector $\mathbf{p}(t_0)$, and the number of jumps is $n_x=n_y=0$. The joint PDF that the particle dwells in a specific site  and that the current has a certain value is thus initially given by $\bar{\mathbf{p}}(\mathbf{n},t_0,t_0) = \mathbf{p}(t_0) \delta(\mathbf{n})$. After time $dt$, the number of jumps attains nonzero values solely by jumps described by the transition rates 
\begin{itemize}
\item $\!r^{i_{y_a}}_{i_{x_a}\to i_{x_a} + 1}(t_0)$,\quad $n_x$ increases by~$1$,
\item $\!r^{i_{y_a}}_{i_{x_a} + 1\to i_{x_a}}(t_0)$,\quad $n_x$ decreases by~$1$,
\item $\!l^{i_{x_a}}_{i_{y_a}\to i_{y_a} + 1}(t_0)$,\quad $n_y$ increases by $1$,
\item $\!l^{i_{x_a}}_{i_{y_a}+1\to i_{y_a}}(t_0)$,\quad $n_y$ decreases by $1$.
\end{itemize}
Using the Master equation~\eqref{eq:MasterMatr}, the vector of occupation probabilities at time $t_0+dt$ can, for short $dt$, be written  as $\mathbf{p}(t_0+dt) = \mathcal{U}(t_0+dt,t_0)\mathbf{p}(t_0)$, where $\mathcal{U}(t_0+dt,t_0) = \left[{\mathcal I} + dt{\mathcal R}(t_0) \right]$ and ${\mathcal I}$ denotes the identity matrix. The joint PDF for the dimensionless current and position at time $t_0+dt$ can be written as 
\begin{equation}
\bar{\mathbf{p}}(\mathbf{n},t_0+dt,t_0) = \bar{\mathcal{U}}(t_0+dt,t_0,\mathbf{n})\mathbf{p}(t_0)\;.
\end{equation}
Here, all the matrix elements of $\bar{\mathcal{U}}(t_0+dt,t_0,\mathbf{n})$ are given by the matrix elements of $\mathcal{U}(t_0+dt,t_0)$ which are multiplied by $\delta(n_x)\delta(n_y)$ except for the four elements containing the above mentioned transition rates. The corresponding non-vanishing currents $n_{x,y}\neq 0$ are represented by shifted $\delta$-functions. For example, the element of $\mathcal{U}$ containing the rate $r^{i_{y_a}}_{i_{x_a}\to i_{x_a} + 1}$ is in $\bar{\mathcal{U}}$ multiplied by $\delta(n_x - 1)$, the element of $\mathcal{U}$ containing the rate $r^{i_{y_a}}_{i_{x_a} + 1\to i_{x_a}}$ is in $\bar{\mathcal{U}}$ multiplied by $\delta(n_x + 1)$, and similarly for the other two elements.

Using the definition \eqref{eq:mapping}--\eqref{eq:transy} of $\mathbf{p}(t_0)$, the matrix element $\left[\bar{\mathcal{U}}(t_0+dt,t_0,\mathbf{n})\right]_{m n}dn_x dn_y$ stands for the joint probability that a particle starting at time $t_0$ from site $[i_x(n),i_y(n)]$ will arrive to site $[i_x(m),i_y(m)]$ at time $t+dt$ given that the numbers of jumps $n_x$ and $n_y$ at site $[i_{x_a},i_{y_a}]$ during the interval $[t_0,t_0+dt]$ assume values from the intervals $(n_x,n_x + dn_x)$ and $(n_y,n_y + dn_y)$. The matrix $\bar{\mathcal{U}}(t_0+2dt,t_0+dt,\mathbf{n})$ allows us to construct the joint PDF $\bar{\mathbf{p}}(\mathbf{n},t_0+2dt,t_0)$ from $\bar{\mathbf{p}}(\mathbf{n},t_0+dt,t_0)$ in a similar manner as $\bar{\mathbf{p}}(\mathbf{n},t_0+ dt)$ from $\mathbf{p}(t_0)$. The only difference is that now the distribution for $\mathbf{n}$ is more involved. Namely, to get the PDF for the current at time $t_0+2dt$, we need to integrate over all possible combinations of the initial $\mathbf{n}$ and the increase in $\mathbf{n}$ during the time interval $dt$: $\bar{\mathbf{p}}(\mathbf{n},t_0+2dt,t_0) = \int dn_x'\int dn_y'\, \bar{\mathcal{U}}(t_0+2dt,t_0+dt,\mathbf{n}')\bar{\mathbf{p}}(\mathbf{n}-\mathbf{n}',t_0+dt,t_0) = [\bar{\mathcal{U}}(t_0+2dt,t_0+dt)\star\bar{\mathbf{p}}(t_0+dt,t_0)](\mathbf{n}) = [\bar{\mathcal{U}}(t_0+2dt,t_0+dt)\star\bar{\mathcal{U}}(t_0+dt,t_0)](\mathbf{n})\mathbf{p}(t_0)$, where $\star$ denotes convolutions in $n_x$ and $n_y$. In a similar manner, one can construct the joint PDF $\bar{\mathbf{p}}(\mathbf{n},t_0+\tau,t_0)$ for the whole time interval $(t_0,t_0+\tau)$. The obvious technical difficulty here lies in the fact than such a PDF would contain many convolutions.  

To circumvent this issue it is advantageous to focus on moment generating functions instead of PDFs. According to the definition \eqref{eq:MGF_def}, the MGF is a Laplace transform of the PDF, which transforms convolutions of original functions into products of transformed functions. The joint PDF $\bar{\mathbf{p}}(\mathbf{n},t_0+2dt,t_0)$ is thus transformed in $\mathbf{p}_{\mathbf{s}}(\mathbf{s}_{\mathbf{n}},t_0+2dt,t_0) = \tilde{\mathcal{U}}(t_0+2dt,t_0+dt,\mathbf{s}_{\mathbf{n}})\tilde{\mathcal{U}}(t_0+dt,t_0,\mathbf{s}_{\mathbf{n}})\mathbf{p}(t_0)$, where the matrices $\tilde{\mathcal{U}}(t+dt,t,\mathbf{s}_{\mathbf{n}})$ are given by Laplace transform of the matrices $\bar{\mathcal{U}}(t+dt,t,\mathbf{n})$. These matrices are called \emph{tilted} matrices and they are identical to $\bar{\mathcal{U}}(t+dt,t,\mathbf{n})$ except for the $\delta$-functions $\delta(n_x \mp 1)$ and $\delta(n_y \mp 1)$ in $\bar{\mathcal{U}}(t+dt,t,\mathbf{n})$ that are transformed to the exponentials $\exp(\mp s_{n_x})$ and $\exp(\mp s_{n_y})$ and the $\delta$-functions $\delta(n_x)$ and $\delta(n_y)$ that are both transformed to 1. The vector $\mathbf{p}_{\mathbf{s}}(\mathbf{s}_{\mathbf{n}},t,t_0)$ thus obeys a similar dynamical equation as the probability vector $\mathbf{p}(t_0)$:
\begin{equation}
\frac{d}{d t} \mathbf{p}_{\mathbf{s}}(\mathbf{s}_{\mathbf{n}},t,t_0) = \tilde{\mathcal{R}}_{\mathbf{s}_{\mathbf{n}}}(t) \mathbf{p}_{\mathbf{s}}(\mathbf{s}_{\mathbf{n}},t,t_0)\;,
\label{eq:ME_current_MGF}
\end{equation}
where the \emph{tilted} rate matrix $\tilde{\mathcal{R}}_{\mathbf{s}_{\mathbf{n}}}(t) = \left[\tilde{\mathcal{U}}(t+dt,t,\mathbf{s}_{\mathbf{n}}) - {\mathcal I}\right]/dt$ can be obtained from the rate matrix ${\mathcal{R}}(t)$ multiplying the rate  $r^{i_{y_a}}_{i_{x_a}\to i_{x_a} + 1}(t)$ by $\exp(- s_{n_x})$, the rate $r^{i_{y_a}}_{i_{x_a} + 1\to i_{x_a}}(t)$ by $\exp( s_{n_x})$, the rate by $l^{i_{x_a}}_{i_{y_a}\to i_{y_a} + 1}(t)$ by $\exp(- s_{n_y})$, the rate $l^{i_{x_a}}_{i_{y_a} + 1\to i_{y_a}}(t)$ by $\exp(s_{n_y})$, and keeping all other rates unchanged. For a given $\mathbf{s}_{\mathbf{n}}$, the formula \eqref{eq:ME_current_MGF} can be solved in a similar manner as the formula for \eqref{eq:MasterMatr} for $\mathbf{p}(t)$. For a time-independent tilted rate matrix $\tilde{\mathcal{R}}_{\mathbf{s}_{\mathbf{n}}}$ the solution to Eq.~\eqref{eq:ME_current_MGF} is given by a matrix exponential
\begin{equation}
\mathbf{p}_{\mathbf{s}}(\mathbf{s}_{\mathbf{n}},t,t_0) =
\exp\left[ \tilde{\mathcal{R}}_{\mathbf{s}_{\mathbf{n}}} (t - t_0) \right]
\mathbf{p}(t_0)\;,
\label{eq:charf_ti}
\end{equation}
while for a time-dependent rate matrix $\tilde{\mathcal{R}}_{\mathbf{s}_{\mathbf{n}}}(t)$ the solution should be constructed using the time discretization analogous to the one used in Eq.~\eqref{eq:GreenApprox} with $\Delta_t = (t_0 + t)/N_t$. We get
\begin{equation}
\mathbf{p}_{\mathbf{s}}(\mathbf{s}_{\mathbf{n}},t,t_0) =
\lim_{\Delta_t \to 0} \prod_{i=0}^{i_t(t)}
\tilde{{\mathcal U}}_i(\mathbf{s}_{\mathbf{n}})\mathbf{p}(t_0)\;,
\label{eq:charf_td}
\end{equation}
where $\tilde{{\mathcal U}}_i(\mathbf{s}_{\mathbf{n}}) = \exp\left[ \tilde{\mathcal{R}}_{\mathbf{s}_{\mathbf{n}}}(t_0 + i \Delta_t) \Delta_t \right]$ if $i \ge 1$ and $\tilde{{\mathcal U}}_0 = {\mathcal I}$. The vectors \eqref{eq:charf_ti} and \eqref{eq:charf_td} give moment generating functions for $n_x$ and $n_y$ conditioned on the final state of the system during the evolution. The unconditioned generating function is thus obtained by summing over all final states:
\begin{equation}
\chi(\mathbf{s}_{\mathbf{n}},t,t_0) =  \mathbf{p}^\top_+ \cdot \mathbf{p}_{\mathbf{s}}(\mathbf{s}_{\mathbf{n}},t,t_0)\;.
\label{eq:char_fun_n}
\end{equation}
where $\mathbf{p}^\top_+$ is a vector of ones.

For fine discretizations, the moment generating function $\chi_{\mathbf{n}}(\mathbf{s}_{\mathbf{n}},t_0+\tau,t_0) = \chi_{\mathbf{n}}(s_{n_x},s_{n_y})$ finally approximates the MGF $\chi_{\bar{\mathfrak{J}}}(\mathbf{s}_{\bar{\mathfrak{J}}},t_0+\tau,t_0)=
\chi_{\bar{\mathfrak{J}}}(s_{\bar{\mathfrak{J}}_x},s_{\bar{\mathfrak{J}}_y})$ for the time-averaged current:
\begin{equation}
\chi_{\bar{\mathfrak{J}}}(s_{\bar{\mathfrak{J}}_x},s_{\bar{\mathfrak{J}}_y}) = \lim_{\Delta_x \to 0, \Delta_y \to 0}
\chi_{\mathbf{n}} \left( \frac{s_{n_x}}{\tau \Delta_y},\frac{s_{n_y}}{\tau \Delta_x}\right)\;.
\label{eq:char_complete}
\end{equation}

\begin{acknowledgments}
We thank M. {\v Z}onda and H. Touchette for valuable
comments on a preliminary version of the paper. We also thank the two anonymous referees whose detailed reports helped us to improve the readability of the manuscript. VH gratefully acknowledges support by the Humboldt foundation and by the Czech Science Foundation (project No. 17-06716S). S.S. acknowledges funding by International Max Planck Research Schools ({IMPRS}).
\end{acknowledgments}

\bibliography{HE_ref}

\end{document}